\documentclass[fleqn,usenatbib]{mnras}
\usepackage{orcidlink}
\usepackage{newtxtext,newtxmath}
\usepackage{multirow}
\usepackage[T1]{fontenc}
\usepackage{ulem}
\DeclareRobustCommand{\VAN}[3]{#2}
\let\VANthebibliography\thebibliography
\def\thebibliography{\DeclareRobustCommand{\VAN}[3]{##3}\VANthebibliography}
\usepackage{graphicx}	
\usepackage{amsmath}	
\title[GRB X-ray plateaus from stratified outflows]{The hydrodynamics of stratified ultra-relativistic outflows and the origin of GRB X-ray plateaus}
\author[G. Sadeh et al.]{
Gilad Sadeh$^{1,\orcidlink{0009-0003-0141-6171}}$\thanks{E-mail: gilad.sadeh@aei.mpg.de},
Kenta Hotokezaka$^{1,2}$,
and Masaru Shibata$^{1,3,\orcidlink{0000-0002-4979-5671}}$
\\
$^{1}$\it{Max Planck Institute for Gravitational Physics (Albert Einstein Institute), Am M\"uhlenberg 1,
Potsdam-Golm, 14476, Germany}\\
$^{2}$\it{Research Center for the Early Universe, Graduate School of Science, The University of Tokyo, Bunkyo, Tokyo 113-0033, Japan}\\
$^{3}$\it{Center for Gravitational Physics and Quantum Information, Yukawa Institute for Theoretical Physics, Kyoto University, Kyoto, 606-8502, Japan}
}
\date{Accepted XXX. Received YYY; in original form ZZZ}
\pubyear{\the\year{}}
\begin{document}
\label{firstpage}
\pagerange{\pageref{firstpage}--\pageref{lastpage}}
\maketitle
\begin{abstract}
The origin of the X-ray plateau phase observed in a large fraction of gamma-ray burst afterglows remains debated. We present a novel analytic framework for the hydrodynamics of ultra-relativistic, radially stratified outflows interacting with an external medium. By explicitly accounting for a continuous distribution of Lorentz factors within the ejecta, we derive analytic expressions describing the evolution of a long-lived, mildly relativistic reverse shock and determine its crossing time. Then, we compute the resulting synchrotron emission from both the forward and reverse shocks. The forward shock naturally produces a shallow, long-lasting X-ray decay consistent with the observed properties of X-ray plateaus, including the Dainotti relation, without requiring prolonged central-engine activity or an additional high-energy emission component. We further show that reproducing the observed plateau durations requires a broad distribution of ejecta Lorentz factors, extending down to $\gamma_\text{min}\sim70-100$, consistent with the ultra-relativistic outflow that powers the prompt $\gamma$-ray emission. The reverse shock generates a long-lived millimeter emission component that outshines the forward shock emission at these wavelengths. Both the plateau and reverse shock emission terminate smoothly once the slowest ejecta are processed, marking a transition to the standard Blandford-McKee self-similar evolution. Such stratified outflows are expected on physical grounds, as the ultra-relativistic ejecta responsible for the prompt $\gamma$-ray emission are unlikely to be launched with a single Lorentz factor. This model provides a unified picture in which the same outflow powers the prompt emission, the X-ray plateau, and the subsequent afterglow evolution.
\end{abstract}
\section{Introduction}
Gamma-ray bursts (GRBs) are the most luminous transient events in the Universe, powered by ultra-relativistic (UR) outflows launched during the collapse of massive stars or the merger of compact objects \citep[see][for reviews]{piran_physics_2005,meszaros_gamma-ray_2006,waxman_gamma-ray_2006,nakar_short-hard_2007,berger_short-duration_2014,kumar_physics_2015}. While the prompt $\gamma$-ray emission encodes the physics of internal dissipation within the outflow, the long-lasting afterglow is produced when the relativistic blast wave interacts with the external medium and accelerates non-thermal electrons that radiate via synchrotron emission \citep{meszaros_optical_1997,waxman_gamma-ray--burst_1997,sari_spectra_1998,granot_shape_2002}. The pre-Swift \citep{gehrels_swift_2004,burrows_swift_2005} standard model predicted a relatively smooth, declining afterglow light curve shaped by the deceleration of a relatively uniform relativistic blast wave.
The early time X-ray observations provided by the Swift X-Ray Telescope revealed a far richer phenomenology than expected \citep{nousek_evidence_2006,liang_comprehensive_2007,troja_swift_2007}. A large fraction of GRB afterglows exhibit an extended phase of unusually shallow flux decay, often lasting $10^3-10^5$ seconds, before transitioning to the familiar steeper power-law decline. This feature, commonly referred to as the plateau phase, typically exhibits temporal slopes, $\alpha$, between $0.3\lesssim\alpha\lesssim0.6$, with little to no spectral evolution across the break \citep{liang_comprehensive_2007,ronchini_combined_2023,li_statistical_2026}.
The physical origin of the plateau remains uncertain. Several mechanisms have been proposed. A long-lived energy injection into the external shock is one of the most common explanations. This can be powered either by fallback accretion, by spin-down of a newly formed magnetar, or by the slower portion of an outflow with a broad, and relatively low ($\sim$tens), initial Lorentz factor (LF) distribution, capable of supplying a sustained luminosity that gradually declines over time \citep{zhang_gamma-ray_2002,nousek_evidence_2006,granot_distribution_2006,yu_shallow_2007,dallosso_gamma-ray_2011,metzger_protomagnetar_2011,matsumoto_linking_2020}. Such an injection alters the dynamics of the forward shock (FS), producing a flatter temporal decay without modifying the underlying synchrotron spectrum. However, this standard refreshed-shock interpretation encounters significant difficulties.
The shallow decay and long duration of the plateau imply that the energy injected at late times, either by continued central-engine activity or by slower ejecta, substantially exceeds the initial kinetic energy of the fast component responsible for the prompt $\gamma$-ray emission. As a result, evaluating the energy budget at the end of the plateau should naturally lead to an inferred radiative efficiency well below the canonical $\eta \sim 10\%$, in tension with observational constraints \citep{wygoda_energy_2016,beniamini_observational_2019}, whereas comparing the prompt $\gamma$-ray output to the kinetic energy available immediately after the prompt phase, instead requires an extremely high prompt efficiency \citep{ioka_efficiency_2006} far above the efficiencies typically expected from internal-shock models \citep{kobayashi_ultraefficient_2001}.
Several explanations attribute the X-ray plateau to geometric effects rather than to the dynamics of the external shock. One possibility is that the plateau represents high-latitude emission (HLE) from the prompt phase, arising from angular structure in the prompt-emitting region and producing an extended tail of the prompt radiation \citep{oganesyan_structured_2020}. Another class of geometric models invokes off-axis viewing of structured jets, in which the early afterglow is flattened as progressively more energetic regions of the jet become visible to the observer \citep{eichler_case_2006,beniamini_x-ray_2020}. Both interpretations generically predict weaker and softer prompt emission for plateau GRBs; observationally, however, GRBs with X-ray plateaus show prompt energetics comparable to those of GRBs without plateaus. HLE is intrinsically tied to the prompt phase and therefore does not naturally explain the subsequent transition to a FS-dominated afterglow, whereas most observed plateaus evolve smoothly into a standard external-shock decay with little or no spectral evolution across the break. Off-axis structured-jet models, on the other hand, can reproduce plateau-like light curves only for specific combinations of jet structure and viewing angle, thereby placing stringent constraints on the parameter space. These considerations suggest that geometric effects alone are unlikely to provide a generic explanation for the majority of X-ray plateaus.
Another proposed explanation associates the plateau phase with the coasting stage of the relativistic outflow, prior to significant blast wave deceleration. In this scenario, a shallow decay can arise naturally if the jet propagates in a wind-like environment and has a relatively low initial LF, $\sim40$, which extends the coasting phase and produces a light curve with a temporal slope comparable to that observed during the plateau \citep{lei_shallow_2011,shen_coasting_2011,duffell_engine_2015,dereli-begue_wind_2022}. This mechanism does not operate in a constant-density interstellar medium (ISM) environment, where the emission during the coasting phase rises in time. Moreover, reproducing plateaus in this framework requires an extremely extended wind medium $\sim10^{18}$cm, and that the entire outflow has such low LFs, which is in tension with prompt-emission constraints for typical long GRBs \citep{lithwick_lower_2001}.
More complex scenarios include evolving shock microphysics \citep{granot_diagnosing_2006,ioka_efficiency_2006,panaitescu_evidence_2006}, reverse shock (RS)-dominated emission \citep{uhm_mechanism_2007,genet_can_2007}, prior emission \citep{yamazaki_prior_2009}, a thick layer with relativistic RS
\citep{leventis_plateau_2014,kusafuka_ejecta_2025} or a thin layer with a Newtonian RS \citep{kobayashi_onset_2007}.
The interaction of relativistic GRB ejecta with the external medium is commonly described within the forward-reverse shock framework. In this picture, an UR outflow drives a FS into the ambient medium while a RS propagates back into the ejecta, with the dynamics governed by the shell deceleration scale and by whether the RS is relativistic or Newtonian \citep{sari_hydrodynamic_1995}. The subsequent evolution of the FS is well described by the self-similar Blandford-McKee (BM) solution \citep{blandford_fluid_1976}, which, together with the forward-reverse shock system, forms the basis of the standard theoretical description of GRB afterglow dynamics.
Despite its success, the canonical picture relies on idealized assumptions about the structure of the ejecta, most notably the treatment of the outflow as a shell characterized by a single LF and a sharp radial profile.
It is physically expected that relativistic ejecta produced in GRBs are not characterized by a single LF, but instead span a finite distribution of velocities \citep{rees_refreshed_1998,sari_impulsive_2000,granot_distribution_2006,laskar_energy_2015}. Even if the central engine launches material with a complicated time history, differential expansion quickly leads to an effective radial stratification: faster portions of the outflow propagate to larger radii, while slower portions lag behind. Such a structure may arise from intrinsic variability of the engine and/or internal dissipation within the flow, including internal shocks that redistribute energy and momentum. As the ejecta drive a FS into the external medium and the shocked region decelerates, progressively slower material can catch up and transfer its energy to the blast wave. The effective energy of the external shock increases over time, leading to a more gradual deceleration and a correspondingly shallower afterglow decay. Such stratification can significantly modify the dynamics of the RS, making it long-lived and mildly relativistic. Once the remaining energy in the slowest material becomes small compared to the accumulated blast wave energy, the evolution should approach the standard BM solution \citep{blandford_fluid_1976}. While a small number of plateaus terminate in extremely steep declines that are inconsistent with typical external-shock decline, $1\lesssim\alpha\lesssim1.5$ \citep{zhang_physical_2006,liang_comprehensive_2007,troja_swift_2007,ror_investigating_2025,swain_grb_2025}, such cases are not common. The majority of observed X-ray plateaus appear broadly compatible with interpretations based on the dynamics and structure of the relativistic outflow \citep{ronchini_combined_2023}.
Several empirical observations further motivate considering GRB outflows with a broad LF distribution. Some short GRBs display extended emission \citep{norris_short_2006}, i.e., a softer and less variable component lasting $\sim10$--$100\,$s, and a subset of these events exhibit X-ray plateaus \citep{gompertz_magnetar_2014,kisaka_long-lasting_2015,matsumoto_linking_2020}. In addition, Einstein Probe soft X-ray monitoring has revealed prompt activity extending beyond the $\gamma$-ray $T_{90}$ \citep{li_minutes-long_2026,fraija_grb250704bep250704a_2026}, followed by an X-ray plateau, demonstrating that prompt emission can persist at softer energies and later times than inferred from $\gamma$-ray data alone. While these observations do not uniquely determine the duration of central-engine activity, they motivate modeling the ultra-relativistic ejecta as radially stratified, rather than as a single-LF shell.
In this work, we revisit the hydrodynamics of GRB outflows interacting with an external medium, extending the standard single LF shell description to radially stratified ejecta \citep{sadeh_nonthermal_2025}. Adopting a quasi-static analytic approach, we derive compact expressions for the shock dynamics, recover the canonical results in the appropriate limits, and develop an analytic description of the transition between relativistic and Newtonian RSs. We then apply this framework to provide a physical interpretation of the GRB X-ray plateau, in which the UR outflow responsible for the prompt emission possesses a broad distribution of LFs, extending down to $\gamma_\text{min}\sim70-100$. In this picture, a mildly relativistic long-lived RS propagates through the ejecta, gradually transferring energy to the shocked region over a timescale set by the mass and minimum LF of the stratified outflow. The plateau ends once the RS completes its passage through the ejecta, after which the blast wave transitions to the standard BM self-similar evolution smoothly. This model directly links the plateau duration to physical properties of the outflow and provides a unified interpretation of the prompt emission, the X-ray plateau, and the subsequent afterglow evolution, without invoking prolonged central-engine activity, finely tuned angular jet structures, or uniformly low LFs of $\sim30$.
The paper is organized as follows. In \S~\ref{sec:hydro}, we develop an analytic framework for the hydrodynamics of radially stratified UR outflows interacting with an external medium and derive the resulting forward- and reverse-shock evolution and duration. In \S~\ref{sec:emission}, we compute the synchrotron break frequencies and emission from both the FS and RS under standard microphysical assumptions. In \S~\ref{sec:obs}, we apply the stratified-outflow model to the observed properties of X-ray plateaus, deriving constraints on the ejecta stratification and the allowed parameter space, and discuss the implications of these constraints for the associated broadband emission. Finally, our conclusions are summarized in \S~\ref{sec:conclusions}.
\section{Forward-Reverse Shock Dynamics}
\label{sec:hydro}
\subsection{Analytic expressions for the shocked plasma LFs}
\begin{figure}
\includegraphics[width=\columnwidth]{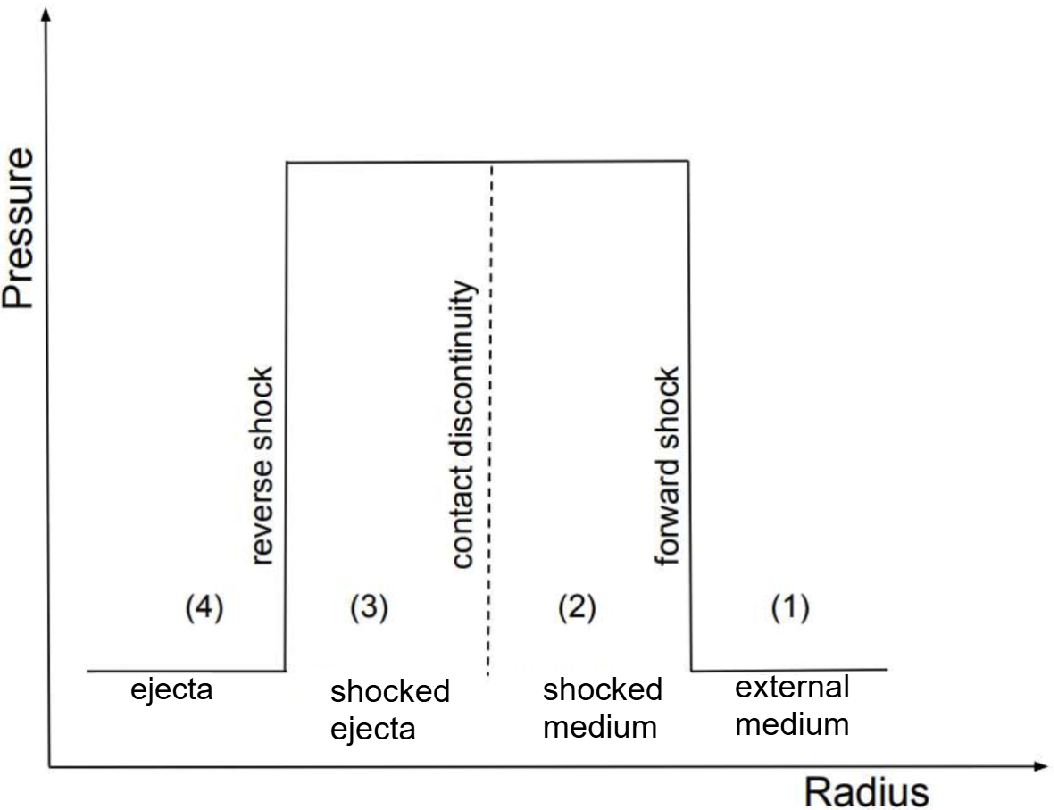}
    \caption{A schematic illustration of the forward-reverse shock structure, showing the four dynamical regions: (1) unshocked external medium, (2) shocked external medium, (3) shocked ejecta, and (4) unshocked ejecta. The pressure in the unshocked regions is negligible compared to that in the shocked regions, $p_1,p_4\ll p_2,p_3$. In our analytic treatment, the pressure and velocity are approximated as uniform throughout the shocked region, while the density is taken to be uniform within each of the regions (2) and (3), separated by the contact discontinuity.}
    \label{fig:FRshocks}
\end{figure}

\begin{figure}
\includegraphics[width=\columnwidth]{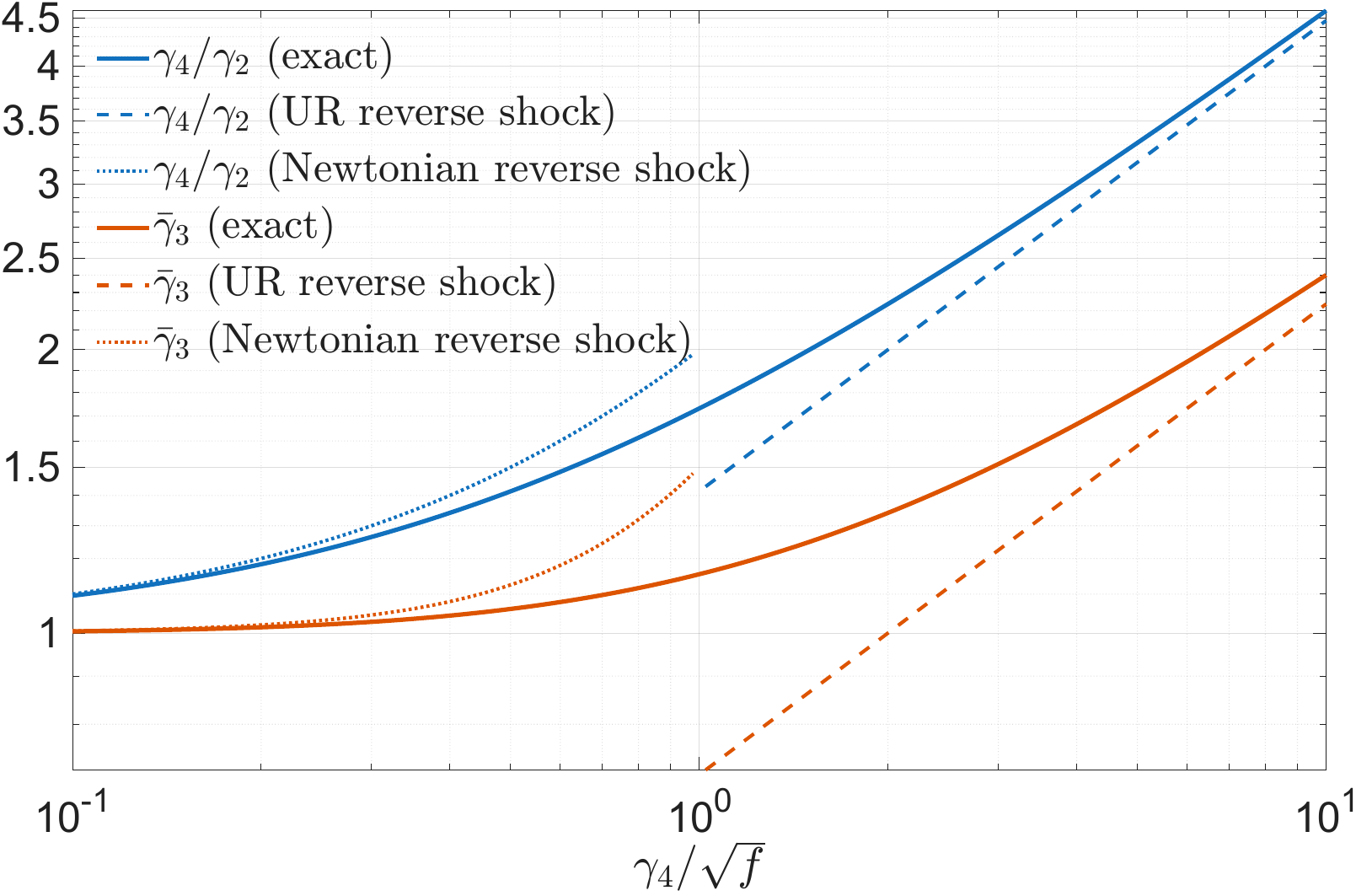}
    \caption{
Exact analytic solutions for the LF ratio, $\gamma_4/\gamma_2$, and for the shocked-ejecta LF, $\bar{\gamma}_3$ (measured in the unshocked ejecta frame), shown as functions of the single dimensionless parameter $\gamma_4/\sqrt{f}$ (with $f\equiv\rho_4/\rho_1$). Also shown are the asymptotic limits: the UR RS regime ($\gamma_4^2\gg f$, Eq.~(\ref{eq:LorUR})) and the Newtonian RS regime ($\gamma_4^2\ll f$, Eq.~(\ref{eq:LorNew})).}
    \label{fig:gamma23}
\end{figure}

We first summarize the solution of the problem in which a shell of UR cold dense matter with a LF $\gamma_4\gg1$ collides with a stationary cold external medium. Two shocks form: a FS that propagates into the external medium, and a RS that propagates into the shell (see Fig. \ref{fig:FRshocks} for a schematic illustration).
The deceleration imposed by the RS reduces kinetic energy in the ejecta, which is subsequently transmitted to the heated ionized plasma within the shocked external medium. While the FS is UR (assuming the density of the shell is larger than the external medium density), the RS is not necessarily so. Therefore, careful analysis is required to fully characterize its hydrodynamic evolution.
We approximate the shocked layers behind the shocks with uniform flow profiles ($e,\rho$ and $\gamma$), given by the shock jump conditions \citep{blandford_fluid_1976},
\begin{equation}
\label{eq:jump0}
    \begin{aligned}
    \rho'=\frac{\hat{\gamma}\gamma+1}{\hat{\gamma}-1}\rho,\quad e=(\gamma-1)\rho'c^2,
    \end{aligned}
\end{equation}
where $\rho$ and $\rho'$ are the mass densities of the unshocked and shocked material, respectively, both measured in their local rest frames, and $\gamma$ is the LF of the shocked fluid measured in the unshocked frame. Such an approximation was successfully tested against fully relativistic hydrodynamics numerical calculations of mildly relativistic \citep{sadeh_non-thermal_2023,sadeh_late-time_2024}, highly relativistic \citep{sadeh_nonthermal_2025}, and UR \citep{kobayashi_hydrodynamics_1999,kusafuka_ejecta_2025} ejecta. The adiabatic index $\hat{\gamma}$ relates the internal energy density, $e$, and pressure, $p$, via $p=(\hat{\gamma}-1)e$.
Because the RS may be mildly relativistic or Newtonian and the heating is inefficient, we adopt an effective adiabatic index of the form
\begin{equation}
    \hat{\gamma}=\frac{4+\left(1+\frac{e}{\rho'c^2}\right)^{-1}}{3},
\end{equation}
which smoothly interpolates between $\hat{\gamma}=4/3$ in the UR limit and $\hat{\gamma}=5/3$ in the Newtonian limit, providing a good approximation to the Synge equation of state (EoS) for an electron-proton plasma \citep{synge_relativistic_1957,kumar_evolution_2003,mignone_equation_2007,uhm_semi-analytic_2011,sadeh_late-time_2024}. For this EoS, the shocked density simplifies to
\begin{equation}
     \rho'=\frac{4\gamma/3+4/3}{1/3\gamma+1/3}\rho=4\gamma \rho,
\end{equation}
which is valid in both the UR and Newtonian regimes. Applying these relations to the forward- and reverse-shocked regions yields
\begin{equation}
    \begin{aligned}
    \label{eq:jump}
        \rho_2=&4\gamma_2\rho_1,\\ \quad \rho_3=&4\bar{\gamma}_3\rho_4,\\
        e_2=&(\gamma_2-1)\rho_2c^2\approx4\gamma_2^2\rho_1c^2,\quad \\e_3=&(\bar{\gamma}_3-1)\rho_3c^2=4(\bar{\gamma}_3^2-\bar{\gamma}_3)\rho_4c^2,\\
         p_2=&\frac{e_2}{3}\approx\frac{4}{3}\gamma_2^2\rho_1c^2,\quad \\p_3=&\left(\frac{1+\frac{1}{\bar{\gamma}_3}}{3}\right)e_3=\frac{4}{3}(\bar{\gamma_3}^2-1)\rho_4c^2,
    \end{aligned}
\end{equation}
where the numerical subscripts denote the regions shown in Fig.~\ref{fig:FRshocks}. $\gamma_2$ and $\gamma_4$ are measured in the external medium frame, and $\bar{\gamma}_3=\gamma_2\gamma_4\left(1-\sqrt{1-\frac{1}{\gamma_2^2}-\frac{1}{\gamma_4^2}+\frac{1}{\gamma_2^2\gamma_4^2}}\right)\approx \frac{1}{2}\left(\frac{\gamma_4}{\gamma_2}+\frac{\gamma_2}{\gamma_4}\right)$ is the LF of the shocked ejecta measured in the unshocked ejecta frame, assuming equal velocities across the contact discontinuity.
Imposing  pressure balance across the contact discontinuity, $p_2\simeq p_3$, yields a closed relation between $\gamma_2$, $\gamma_4$, and the density ratio $f\equiv\rho_4/\rho_1$, following the notation of \cite{sari_hydrodynamic_1995},
\begin{equation}
    \begin{aligned}
\gamma_2^2=f[\bar{\gamma_3}^2-1]=f\left[\frac{1}{4}\left(\frac{\gamma_2^2}{\gamma_4^2}+\frac{\gamma_4^2}{\gamma_2^2}\right)-\frac{1}{2}\right],\\
(4\gamma_4^2-f)\gamma_2^4+2f\gamma_4^2\gamma_2^2-f\gamma_4^4=0.
\end{aligned}
\end{equation}
Remarkably, this equation admits a compact analytic solution \citep[see also][]{zhang_physics_2018,zhang_semi-analytical_2022},
\begin{equation}\begin{aligned}
\label{eq:Lor}
\frac{\gamma_4}{\gamma_2}=\sqrt{1+2\frac{\gamma_4}{\sqrt{f}}},\quad \bar{\gamma}_3
    =\frac{1}{2}\left(\frac{1}{\sqrt{1+2\frac{\gamma_4}{\sqrt{f}}}}+\sqrt{1+2\frac{\gamma_4}{\sqrt{f}}}\right).\\
\end{aligned}
\end{equation}
The applicability of this solution requires $\gamma_2\gg1$, which implies $\gamma_4/\sqrt{1+2\gamma_4/\sqrt{f}}\gg1$.
Two limiting regimes are of particular interest. In the UR RS limit, $\gamma_4^2\gg f$, we recover the standard results of \cite{sari_hydrodynamic_1995},
\begin{equation}
    \begin{aligned}
    \label{eq:LorUR}
\gamma_2\approx\frac{f^{\frac{1}{4}}\gamma_4^{\frac{1}{2}}}{\sqrt{2}},\quad \bar{\gamma}_3\approx  \frac{\gamma_4^\frac{1}{2}}{\sqrt{2}f^\frac{1}{4}}.   \end{aligned}
\end{equation}
While in the Newtonian RS limit, $\gamma_4^2\ll f$, the use of a variable EoS leads to
\begin{equation}
\begin{aligned}
\label{eq:LorNew}
\gamma_2\approx\gamma_4\left(1-\frac{\gamma_4}{\sqrt{f}}\right),\quad \bar{\gamma}_3\approx1+\frac{\gamma_4^2}{2f}, \\
    \end{aligned}
\end{equation}
slightly different from the result in \citet{sari_hydrodynamic_1995}. In Fig.~\ref{fig:gamma23}, we show the exact solutions for $\gamma_4/\gamma_2$ and $\bar{\gamma}_3$ as functions of $\gamma_4/\sqrt{f}$, together with their asymptotic UR and Newtonian limits. These relations form the basis for the quasi-static treatment of the RS evolution developed in the following subsection.
\subsection{Stratified structure}
\begin{figure}   \includegraphics[width=\columnwidth]{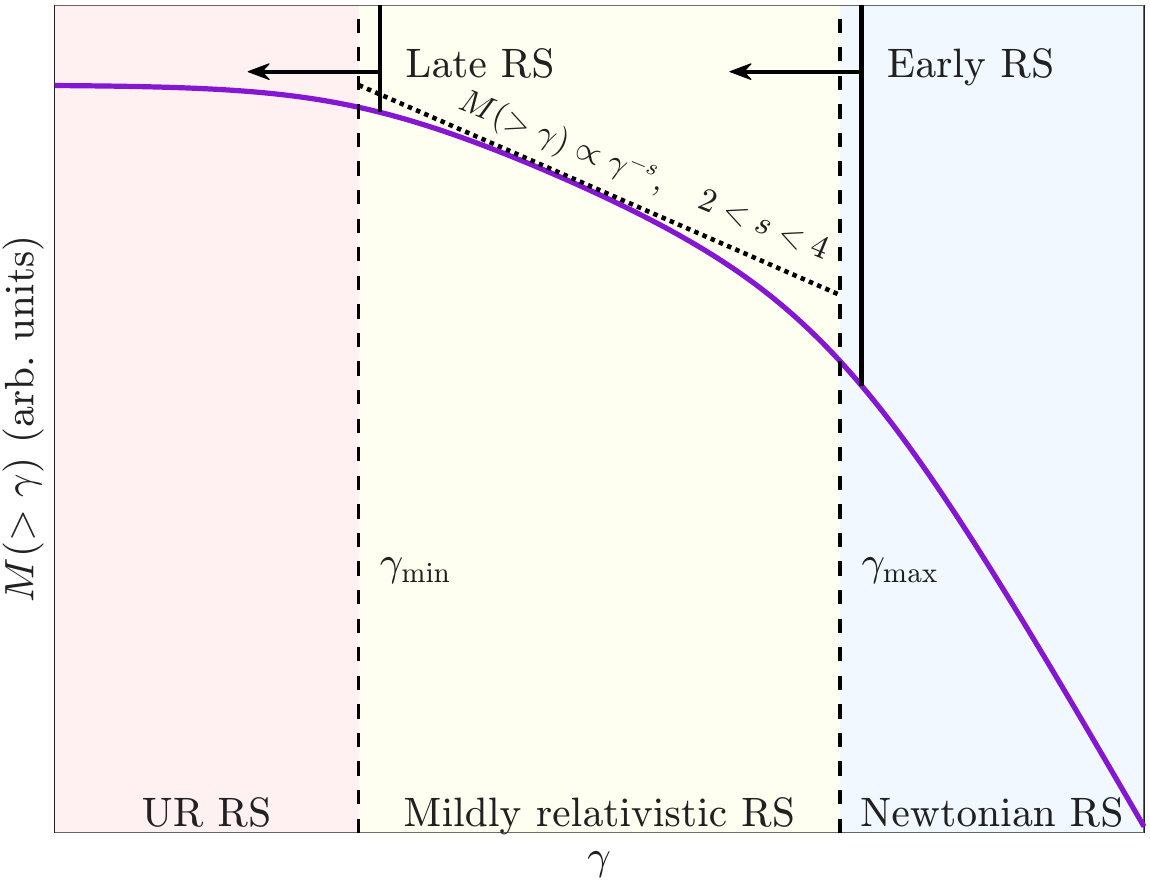}
    \caption{Schematic illustration of the ejecta LF structure. The cumulative mass profile $M(>\gamma)$ is shown as a function of LF, featuring a steep high-$\gamma$ tail above a characteristic LF $\gamma_\text{max}$ and a shallow, approximately constant profile at lower $\gamma_\text{min}$. The RS initially propagates through the steep part of the ejecta and subsequently enters the shallow region, as indicated by the arrows. The figure is qualitative and not drawn to scale.}
    \label{fig:structure}
\end{figure}

\begin{figure}
\includegraphics[width=\columnwidth]{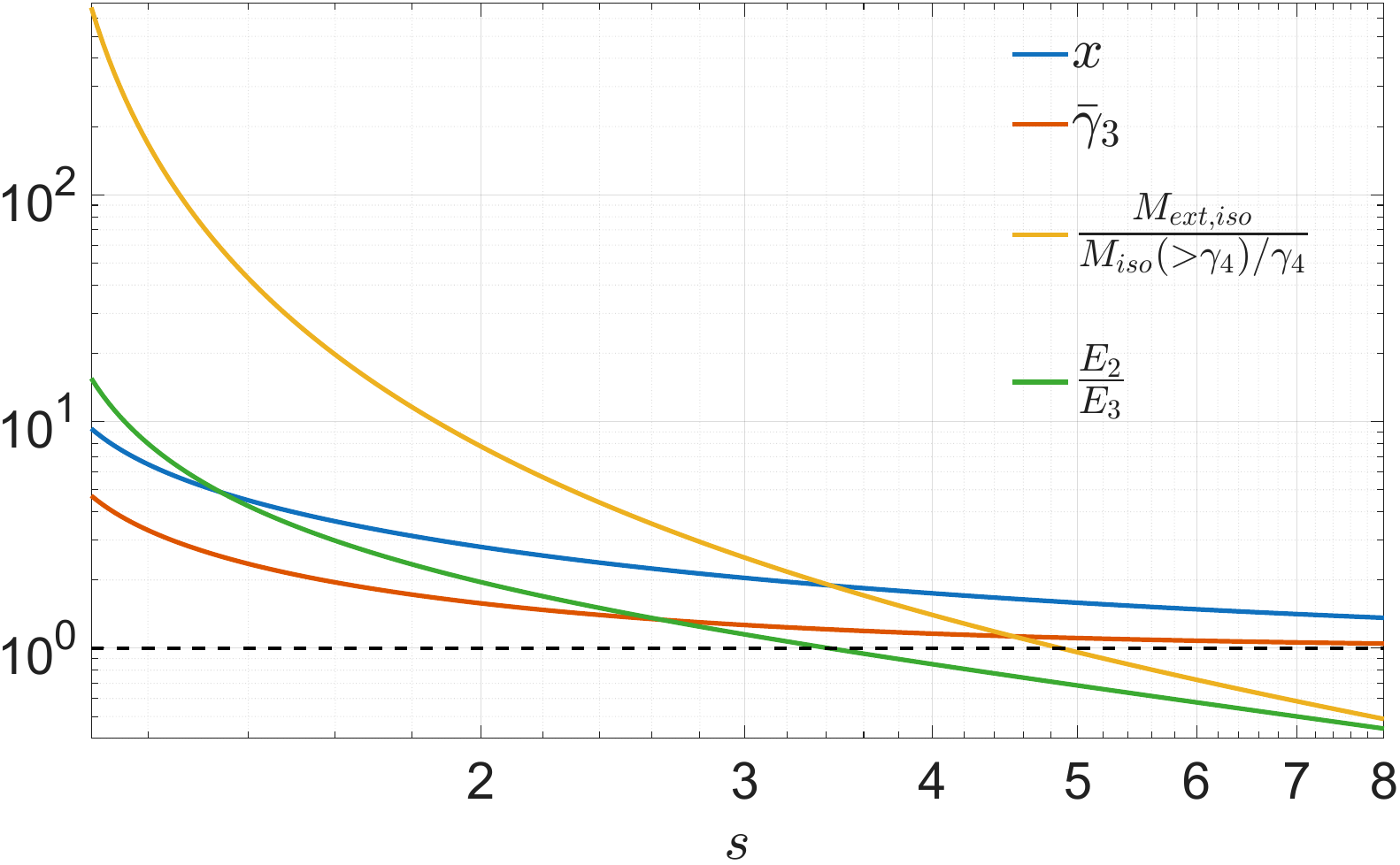}
    \caption{In Blue: numerical solution of Eq.~(\ref{eq:numeric}) for $k=0$ showing the dependence of the LF ratio $x\equiv\gamma_4/\gamma_2$ on the ejecta power-law index $s$. In red: the reverse shock LF in the frame of the unshocked ejecta, $\bar{\gamma}_3=\frac{1}{2}\left(x+\frac{1}{x}\right)$. In yellow: the ratio of the swept-up external medium mass, $M_{\text{ext}}$, to the canonical deceleration mass, $M_\text{iso}(>\gamma_4)/\gamma_4$ \citep{rees_relativistic_1992}. In green: The ratio between the shocked external medium energy, $E_2$, and the shocked ejecta energy, $E_3$. All are shown as a function of the ejecta power-law index $s$, and for $k=0$. For $s\gg1$, corresponding to a Newtonian RS, the outflow experiences negligible deceleration due to the RS; consequently, most of the energy remains in the ejecta ($E_3$).
    }
    \label{fig:xs}
\end{figure}

\begin{figure}  \includegraphics[width=\columnwidth]{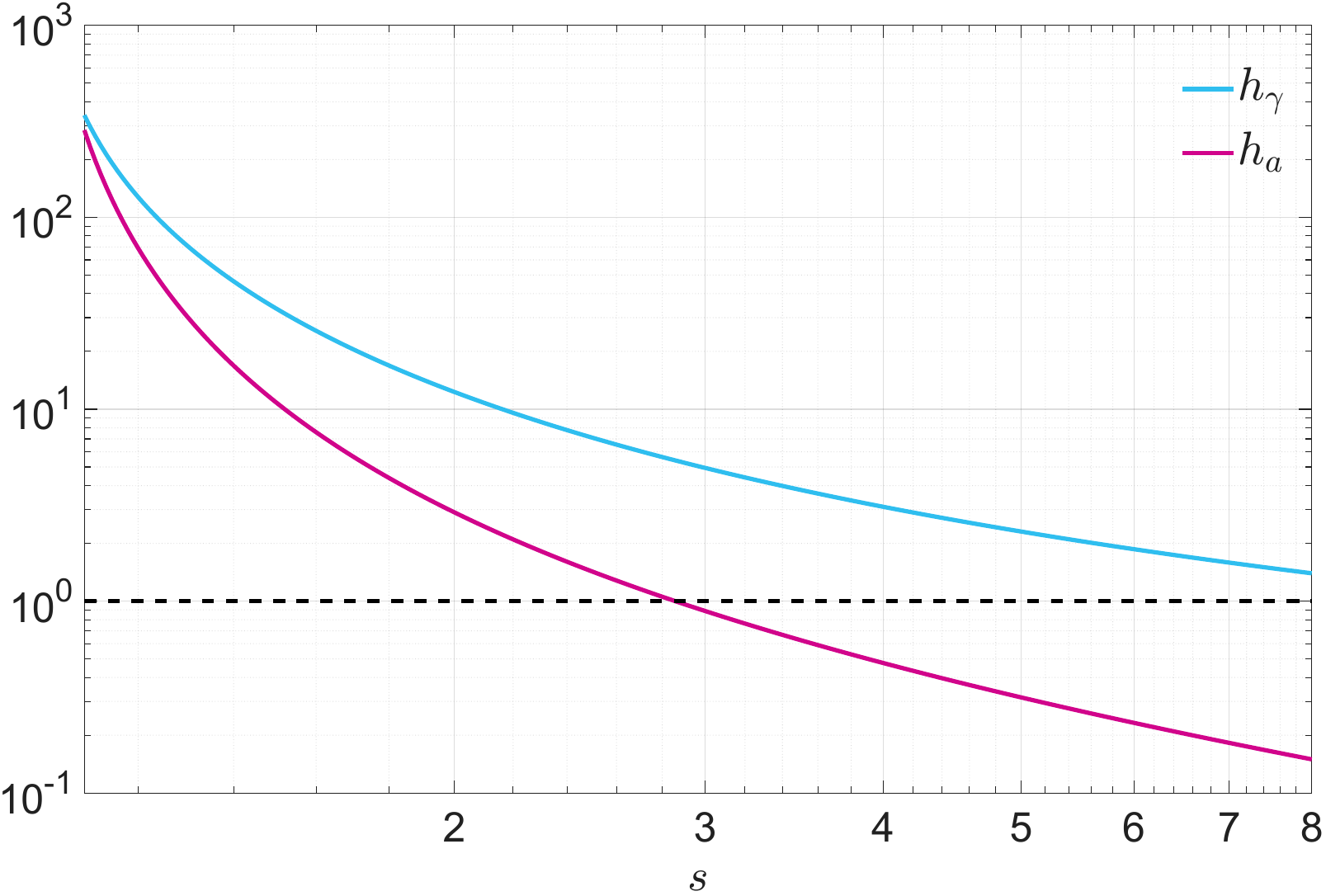}
    \caption{The corresponding correction factors for the RS crossing time, $h_\gamma$ (uniform ISM; $k=0$) and $h_a$ (wind environment; $k=2$), are shown. For moderate values $2\lesssim s\lesssim 4$, relevant when the RS propagates through the shallow part of the ejecta, the RS is mildly relativistic and the crossing time is significantly extended relative to the standard thin-shell estimate in the ISM case.}
    \label{fig:t}
\end{figure}

Throughout this work, we use isotropic-equivalent energies and masses. This approximation is justified by the UR nature of the outflow and the typical opening angles inferred for GRB jets ($\theta_j\sim0.1\gg1/\Gamma$), which ensure that the early-time dynamics and emission are effectively insensitive to lateral expansion and jet-edge effects. As a result, the hydrodynamic evolution during the phases of interest can be accurately described within a spherically symmetric framework.
In general, any relativistic outflow is expected to exhibit a distribution of LFs. Consequently, the cumulative mass above a given cutoff, $M_\text{iso}(>\gamma)$, should decline steeply above some maximum LF $\gamma_\text{max}$, while approaching a constant value below a minimum LF $\gamma_\text{min}$. A schematic illustration of this structure is shown in Fig.~\ref{fig:structure}. The RS initially propagates through the steep high-$\gamma$ tail of the ejecta, before reaching the shallower, low-$\gamma$ region.
Below, we present a quasi-static analysis of relativistic ejecta propagating into a cold external medium, parametrized by a cumulative mass profile $M_\text{iso}(>\gamma)\propto \gamma^{-s}$, valid for $s>1$, so the cumulative energy above any arbitrary low LF reaches a constant \citep[see also][]{sari_impulsive_2000}. The unshocked ejecta is modeled as a sequence of radially stratified shells, each coasting at a LF $\gamma_4$. We assume that a forward-reverse shock structure is established and maintained throughout the evolution. This parametrization provides an adequate description of the ejecta structure across the different stages of the RS propagation, while reproducing the correct temporal scaling of the shocked-plasma LF in the limits corresponding to steep ($s\rightarrow\infty$) and shallow ($s\rightarrow1$) ejecta profiles. For moderate values of $2\lesssim s\lesssim 4$, the characteristic width of the ejecta is given by $\sim\frac{R}{2\gamma_\text{min}^2}$, where $R$ is the FS radius.
    We consider relativistic ejecta characterized by a cumulative mass profile
\begin{equation}
M_\text{iso}(>\gamma_4)=M_\text{iso}\left(\frac{\gamma_4}{\gamma_\text{min}}\right)^{-s}, \qquad s>1.
\end{equation}
A shell with a single LF is recovered in the limit $s\rightarrow\infty$, for which $\gamma_4\rightarrow\gamma_\text{min}$. The isotropic equivalent energy of the ejecta is set by
\begin{equation}
\begin{aligned}
    \label{eq:energy}
E_\text{iso}(>\gamma_4)&=\int\frac{dM_\text{iso}(>\gamma_4)}{d\gamma_4}(\gamma_4-1)c^2d\gamma_4,\\&\approx \frac{s}{s-1}\gamma_4M_\text{iso}(>\gamma_4)c^2\equiv E_\text{iso}\left(\frac{\gamma_4}{\gamma_\text{min}}\right)^{1-s},
\end{aligned}
\end{equation}
where we assumed $\gamma_4\gg1$ and $E_\text{iso}$ is the total isotropic equivalent energy.
The total energy conservation equation follows from conservation of the energy-momentum tensor, $\partial_\mu T^{\mu 0}=0$, such that the conserved quantity can be written as
\begin{equation}
    \int T^{00}dV=\int \left((\varepsilon+p)\gamma^2-p\right)dV=\int \left((\varepsilon+p)\gamma^2-p\right)\frac{dM}{\rho\gamma},
\end{equation}
where $\varepsilon$ includes the rest mass energy. The kinetic energy carried by ejecta with LFs exceeding a given $\gamma_4$ is transferred to the combined energy of the shocked external medium and the shocked ejecta,
\begin{equation}
\label{eq:en_con}
    \begin{aligned}
        E_\text{iso}(>\gamma_4)&= E_2+E_3,
    \end{aligned}
\end{equation}
where $E_{i}$ denotes the total energy in region $i$ (see Fig. \ref{fig:FRshocks}). The total energies in regions (2) and (3) are given by \citep{sadeh_late-time_2024}
\begin{equation}
    \begin{aligned}
        E_2&= \frac{4}{3}\gamma_2^2M_\text{ext,iso}c^2,\\
E_3&=\gamma_2\left(\frac{4\bar{\gamma_3}-\frac{1}{\bar{\gamma_3}}}{3}\right)M_\text{iso}(>\gamma_4)c^2.
    \end{aligned}
\end{equation}
Energy conservation then implies\footnote{Note that the customary pressure-balance prescription does not by itself guarantee global energy conservation, as demonstrated by \citet{uhm_semi-analytic_2011}. We avoid this problem by imposing energy conservation explicitly to derive the blast wave evolution. For full self-similar treatment, see \citet{nakamura_self-similar_2006}. This method was verified against full numerical relativistic hydrodynamical computations in \citet{sadeh_non-thermal_2023}.}
\begin{equation}
        \frac{s}{s-1}\gamma_4M_\text{iso}(>\gamma_4)=\frac{4}{3}\gamma_2^2M_\text{ext,iso}
+\gamma_2\left(\frac{4\bar{\gamma_3}-\frac{1}{\bar{\gamma_3}}}{3}\right)M_\text{iso}(>\gamma_4).
    \end{equation}
    The swept-up external medium mass is therefore
    \begin{equation}
    \begin{aligned}
    \label{eq:Mism}
    M_\text{ext,iso}=%
    \frac{3}{4\gamma_4}M_\text{iso}(>\gamma_4)\left[\frac{s}{s-1}x^2-\frac{2}{3}\left(x^2+\frac{1}{x^2+1}\right)\right],
    \end{aligned}
    \end{equation}
where we have defined $x\equiv\frac{\gamma_4}{\gamma_2}$.
In the UR/Newtonian RS limits, the swept-up external medium mass reduces to
\begin{equation}
M_\text{ext,iso}=
    \begin{cases}
        \frac{3}{2\sqrt{f}}M_\text{iso}(>\gamma_4)\left[\frac{s}{s-1}-\frac{2}{3}\right],\quad \text{UR RS limit,}\\
        \frac{3}{4\gamma_4}M_\text{iso}(>\gamma_4)\left[\frac{s}{s-1}-1\right],\quad \text{Newtonian RS limit}.
    \end{cases}
\end{equation}
The shocked external medium mass, $M_\text{ext,iso}$, and the shocked ejecta mass, $M_\text{iso}(>\gamma_4)$, can be related to their corresponding mass densities as
\begin{equation}
\label{eq:dens}
    \begin{aligned}
        \rho_1&=\frac{(3-k)M_\text{ext,iso}}{4\pi R^3},\quad
        \rho_4&=\frac{\gamma_4 s M_\text{iso}(>\gamma_4)}{4\pi (R_\text{cd}-\Delta_\text{ej})^3},
    \end{aligned}
\end{equation}
where we consider an external density profile $\rho_1 = A r^{-k}$, $R_\text{cd}$ is the contact discontinuity radius and $\Delta_\text{ej}$ the thickness of the shocked ejecta layer.
To estimate the thickness of the shocked external medium layer, $\Delta_\text{ext}$, we impose mass conservation across the FS,
\begin{equation}
\begin{aligned}
    \frac{4\pi}{3-k}\rho_1R^3&=\frac{4\pi}{3}4\gamma_2^2\rho_1\left(R^3-(R-\Delta_\text{ext})^3\right),  \\
    \Delta_\text{ext}&=R\left(1-\frac{\left(\frac{4(3-k)}{3}\gamma_2^2-1\right)^\frac{1}{3}}{\left(\frac{4(3-k)}{3}\gamma_2^2\right)^\frac{1}{3}}\right)\approx\frac{R}{4(3-k)\gamma_2^2},
\end{aligned}
\end{equation}
where the final approximation assumes $\gamma_2\gg1$.
The thickness of the shocked ejecta layer, $\Delta_{\text{ej}}$, follows from mass conservation across the RS,
\begin{equation}
\begin{aligned}
    M_\text{iso}(>\gamma_4)&=\frac{4\pi}{3}4\gamma_2\bar{\gamma}_3\rho_4\left(R_\text{cd}^3-(R_\text{cd}-\Delta_\text{ej})^3\right),  \\
\Delta_\text{ej}&=R_\text{cd}\left(1-\frac{\left(\frac{\gamma_2\gamma_4 s}{3}\cdot4\bar{\gamma}_3\right)^\frac{1}{3}}{\left(1+\frac{\gamma_2\gamma_4 s}{3}\cdot4\bar{\gamma}_3\right)^\frac{1}{3}}\right)
\approx \frac{R_\text{cd}}{ 2s(\gamma_4^2+\gamma_2^2)},
\end{aligned}
\end{equation}
where the final expression applies in the relativistic limit.
Since the contact discontinuity satisfies $R-\Delta_\text{ext}=R_{\text{cd}}$, we obtain
\begin{equation}
\label{eq:r_del}
    (R_\text{cd}-\Delta_\text{ej})^3
    \approx R^3\left(1-\frac{3}{4(3-k)\gamma_2^2}-\frac{3}{ 2s(\gamma_4^2+\gamma_2^2)}\right).
\end{equation}
Finally, the density ratio between the unshocked ejecta and the ambient medium is given by
\begin{equation}
    \begin{aligned}
        f&=\frac{\rho_4}{\rho_1}=\frac{s\gamma_4}{3-k}\frac{M_\text{iso}(>\gamma_4)}{M_\text{ext,iso}}\frac{R^3}{(R_\text{cd}-\Delta_\text{ej})^3}\\&\approx\frac{4\gamma_4^2s}{3(3-k)\left[\frac{s}{s-1}x^2-\frac{2}{3}\left(x^2+\frac{1}{x^2+1}\right)\right]}.
    \end{aligned}
\end{equation}
Using Eq. (\ref{eq:Lor}), the following implicit expression is derived
\begin{equation}
\label{eq:numeric}
    s(x^2-1)^2=3(3-k)\left[\frac{s}{s-1}x^2-\frac{2}{3}\left(x^2+\frac{1}{x^2+1}\right)\right].
\end{equation}
Notably, the ratio $x\equiv\gamma_4/\gamma_2$ depends only on the power-law index $s$ and the external density slope $k$, and is independent of the absolute values of $\gamma_4$ and $\gamma_2$. In Fig.~\ref{fig:xs}, we solve Eq.~(\ref{eq:numeric}) numerically for $k=0$ to obtain $x(s,k=0)$ along with $\bar{\gamma}_3(s,k=0)$. For $k=2$, the smaller coefficient in Eq.~(\ref{eq:numeric}) yields smaller ratios, $x(s,k{=}2)\simeq1.92,\,1.52,\,1.37$ for $s=2,\,3,\,4$, respectively (compared to $x(s,k{=}0)\simeq2.8,\,2.04,\,1.75$). For moderate values of $2\lesssim s\lesssim 4$, as expected when the RS propagates into the shallower part of the ejecta, the RS is mildly relativistic, with $x$ of order a few \citep{waxman_neutrino_2000}. We emphasize that, for any specific choice of $\gamma_4$, the condition $\gamma_2\gg1$ must be satisfied for the present analysis to remain valid.
Eq.~(\ref{eq:Mism}) can be recast in the compact form
\begin{equation}
\label{eq:Mism1}
    M_\text{ext,iso}=
    \frac{M_\text{iso}(>\gamma_4)}{\gamma_4}\cdot\frac{s(x^2-1)^2}{4(3-k)}.
\end{equation}
In Fig.~\ref{fig:xs}, we also plot the ratio of the swept-up external medium mass, $M_\text{ext,iso}$, to the canonical estimate for the external medium mass required for deceleration \citep{rees_relativistic_1992}, $M(>\gamma_4)/\gamma_4$, along with the ratio between the energy in the shocked external medium, $E_2$, to the energy in the shocked ejecta, $E_3$, both as a function of $s$, and for $k=0$. For $s\gg1$ ($x\rightarrow 1$), corresponding to a Newtonian RS, the outflow is not significantly decelerated by the RS.
We now turn to the temporal evolution of the shocked-plasma LF, $\gamma_2(t)$. As long as the FS remains UR, the swept-up external medium mass for an external density profile $\rho_1 = A r^{-k}$, with $k<3$ and $A$ being constant, is
\begin{equation}
\begin{aligned}
\label{eq:Mism2}
    M_\text{ext,iso}=\frac{4\pi}{3-k}AR^{3-k}\approx\frac{4\pi}{3-k}A\left(\frac{2\gamma_2^2ct}{1+z}\right)^{3-k},
\end{aligned}
\end{equation}
where $t$ is the observer time, related to the shock radius by $t=(1+z)R/(\zeta c\gamma_2^2)$, with $z$ as the cosmological redshift. For self-similar BM evolution $\zeta\approx 4$, while for full coasting $\zeta=1$, assuming emission arrives mainly from an angle $\sim\frac{1}{\gamma}$ relative to the line of sight \citep{waxman_angular_1997,sari_observed_1998,panaitescu_rings_1998}. We adopt here $\zeta=2$ as the deceleration is intermediate.
Combining Eqs.~(\ref{eq:energy}), (\ref{eq:Mism1}) and (\ref{eq:Mism2}) yields \citep[same temporal scaling as][]{sari_impulsive_2000}
\begin{equation}
\label{eq:gamma}
    \gamma_2= t^\frac{3-k}{2k-7-s}\left[\frac{16\pi2^{3-k}c^{5-k}x^{1+s}A\gamma_\text{min}^{1-s}}{(1+z)^{3-k}(x^2-1)^2(s-1)E_\text{iso}}\right]^{\frac{1}{2k-7-s}},
\end{equation}
This expression correctly reproduces the relevant asymptotic limits. In the limit $s\rightarrow\infty$, corresponding to a shell with a single LF, the RS has a negligible dynamical effect and $\gamma_2\simeq\mathrm{const}$. Conversely, for $s\rightarrow1$, the RS has already crossed most of the energetic ejecta, the energy from the slower parts adds up only logarithmically, and the flow asymptotically approaches the instantaneous energy injection case, given by self-similar BM solution \citep{blandford_fluid_1976}, for which $\gamma_2\propto t^{(3-k)/(2k-8)}$.
Thus, while the RS propagates through the steep part of the ejecta profile, one has $\gamma_2\simeq\gamma_4\simeq\mathrm{const}$. As the shock encounters progressively shallower ejecta layers, the temporal evolution of $\gamma_2$ transitions smoothly toward the standard BM deceleration law.
The RS crossing time is set by
\begin{equation}
\begin{aligned}
t_\text{cross}=\frac{(1+z)x^2}{2c\gamma_\text{min}^2}\left[\frac{E_\text{iso}}{16\pi Ac^2\gamma_\text{min}^2}\times(s-1)\left(x^2-1\right)^2\right]^{\frac{1}{3-k}}.
\end{aligned}
\end{equation}
For the cases of a uniform ISM ($k=0$) and a wind profile produced by a constant mass-loss rate ($k=2$), this expression reduces to
\begin{equation}
\label{eq:t}
    \begin{aligned}
    t_\text{cross}&=
\begin{cases}t_\gamma\times\left(\frac{x^6}{12}\times(s-1)\left(x^2-1\right)^2\right)^{\frac{1}{3}}\equiv t_\gamma\times h_\gamma,\quad \text{for } k=0,\\
t_a\times\left(\frac{x^2}{9}\times(s-1)\left(x^2-1\right)^2\right)\equiv t_a\times h_a,\quad \text{for } k=2,
    \end{cases}\\
\end{aligned}
\end{equation}
with
    \begin{equation}
\label{eq:tga}
    \begin{aligned}
t_\gamma&\equiv9\times10^2(1+z)E^\frac{1}{3}_{\text{iso},53}n^{-\frac{1}{3}}_{-2}\gamma^{-\frac{8}{3}}_{\text{min},100}\text{ s},\\
t_a&\equiv 3.3\times10^2(1+z)E_{\text{iso},53}A^{-1}_{10}\gamma^{-4}_{\text{min},100}\text{ s},
    \end{aligned}
\end{equation}
where $n$ is the constant number density of the external medium in case $k=0$, $\gamma_{\text{min},100}=\gamma_\text{min}/100$, and for the other parameters we adopt the following notation $q_x=q/10^x$. $t_\gamma$ is the thin-shell deceleration time for a uniform ISM \citep{sari_hydrodynamics_1997,sari_hydrodynamic_1995}, and $t_a$ is the corresponding deceleration time for a wind environment \citep{shen_coasting_2011,dereli-begue_wind_2022}. In Fig.~\ref{fig:t}, we plot the correction factors $h_\gamma$ and $h_a$ as functions of the ejecta power-law index $s$. For moderate values $s\simeq2$, as expected when the RS propagates through the shallow part of the ejecta, the crossing time is increased by approximately an order of magnitude relative to the standard estimate $t_\gamma$. In practice, this implies that when the ejecta has a non-negligible radial width, as expected for an outflow with a distribution of LFs, the transition from the coasting phase to the BM phase is not abrupt. Instead, an intermediate stage emerges in which the deceleration proceeds gradually.
\section{Synchrotron emission from the shocked regions}
\label{sec:emission}
Following the hydrodynamic evolution of the shocked regions, we calculate the synchrotron emission produced by both the FS and RS during the phase in which the RS propagates through the stratified ejecta. We map this dynamical evolution to observable synchrotron emission within the standard afterglow framework, explicitly accounting for the modified dynamics induced by the ejecta stratification. The resulting temporal scalings are consistent with those first derived by \citet{sari_impulsive_2000}. For brevity, in the main text we present normalized expressions only for the two benchmark cases $k=0$ and $k=2$. The corresponding temporal scalings for arbitrary $k$ are summarized in Appendix \ref{app:k}. The observed X-ray emission is dominated by synchrotron radiation from the forward-shocked external medium. The RS primarily contributes at lower frequencies, most notably in the millimeter (mm) band, while its role at high energies is negligible. Its main influence on the X-ray emission is therefore indirect, through its regulation of the energy transfer from the stratified ejecta to the FS.
We adopt the standard assumptions commonly used in GRB afterglow modeling \citep{waxman_gamma-ray--burst_1997,waxman_angular_1997,sari_hydrodynamics_1997,sari_spectra_1998}: Fixed fractions
$\varepsilon_{e}$ and $\varepsilon_{B}$ of the post-shock internal energy density are transferred to relativistic electrons and magnetic fields, respectively. We further assume those fractions are the same for the shocked external medium and the shocked ejecta, $\varepsilon^f_{e/B}=\varepsilon^r_{e/B}$, where $f/r$ notation corresponds to the FS/RS. The electrons are accelerated into a power-law distribution, $N(\gamma_e)\,{ d}\gamma_e \propto \gamma_e^{-p}\,{ d}\gamma_e,
\gamma_e \ge \gamma_m$, with $p\geq2$, and
\begin{equation}
\label{eq:gammam}
    \gamma^f_m = \varepsilon_e\gamma^f\frac{l_p}{p-1}\cdot\frac{m_p}{m_e}\equiv  \varepsilon_e\gamma^f f_p\frac{m_p}{m_e},
\end{equation}
\begin{equation}
\label{eq:reverse1}
    \gamma^r_m =  \varepsilon_e(\gamma^r-1) f_p\frac{m_p}{m_e}=\left(\frac{(x-1)^2}{2x\gamma^f}\right)\gamma_m^f,
\end{equation}
where we simply denote $\gamma_2$ as $\gamma^f$, $\bar{\gamma}_3$ as $\gamma^r$. $l_p$ is given by \citep{sadeh_non-thermal_2023}
\begin{equation}\label{eq:lp}
  l_p=\frac{p-2}{1-(\gamma_\text{M}/\gamma_m)^{2-p}},\quad
  l_p\xrightarrow[p \to 2]{}\frac{1}{\ln(\gamma_\text{M}/\gamma_m)},
\end{equation}
where $\gamma_M\approx\sqrt{3e/(\sigma_T B)}$ is the maximum electron LF in the power law, assuming that the acceleration time equals the synchrotron cooling time \citep{dai_afterglow_2001}.
We further assume that radiative losses are negligible, so that the blast wave evolution remains adiabatic.
This phenomenological description, capturing the post-shock relativistic plasma conditions, finds support across a diverse spectrum of observations and plasma calculations \citep[see][]{blandford_particle_1987,keshet_energy_2005,waxman_gamma-ray_2006,keshet_analytical_2006,spitkovsky_particle_2008,sironi_particle_2009,bykov_fundamentals_2011,sironi_maximum_2013,pohl_pic_2020}.
Following Eq. (\ref{eq:jump}) for the internal energy density behind the FS, the comoving magnetic field there is
\begin{equation}
B^f =(32\pi\varepsilon_B (\gamma^f)^2\rho c^2)^{1/2},
\end{equation}
where we denoted $\rho_1$ as $\rho$.
Pressure equality between the two shocked regions yields the following expression for the comoving magnetic field in the shocked ejecta (using Eq. (\ref{eq:jump}))
\begin{equation}
\label{eq:reverse2}
    B^r=\frac{B^f}{\sqrt{1+\frac{1}{\gamma^r}}}=\frac{\sqrt{x^2+1}}{x+1}B^f.
\end{equation}
The accelerated electrons emit synchrotron radiation. Electrons with a LF above $\gamma_c$ lose a significant part of their energy during the dynamical time. It is given by
\begin{equation}
    \gamma_c= \frac{3\pi m_ec(1+z)}{\sigma_T B^2t\gamma^f}.
\end{equation}
The characteristic synchrotron frequency associated with electrons of LF $\gamma_e$, averaged over an isotropic pitch-angle distribution and over the specific power contribution\footnote{A simple average over the specific power contribution,  $\overline{\nu}=\frac{\int\nu P^\text{syn}_\nu d\nu}{\int P^\text{syn}_\nu d\nu}$.}, $P^\text{syn}_\nu$, is \citep{rybicki_radiative_1979}
\begin{equation}
\label{eq:nusyn}
\nu_\text{syn}(\gamma_e) = \frac{\gamma^f\gamma_e^2eB}{4m_e c(1+z)}.
\end{equation}
Substituting the expressions above and using Eqs (\ref{eq:Lor}) and (\ref{eq:gamma}), the characteristic frequencies corresponding to $\gamma_m$ ($\nu_m\equiv\nu_\text{syn}(\gamma_m)$) and $\gamma_c$ ($\nu_c\equiv\nu_\text{syn}(\gamma_c)$) are
\begin{align}
\nu^f_m &=\begin{cases}5.8\times10^{14}\varepsilon_{e,-1}^2\varepsilon_{B,-2}^{\frac{1}{2}}n_{-2}^{\frac{s-1}{2(s+7)}}E_{\text{iso},53}^\frac{4}{s+7}\gamma_{\text{min},100}^\frac{4(s-1)}{s+7}\times\\\left(f_{s0,2}\right)^4f_{p,-1}^2(1+z)^\frac{5-s}{s+7}t_h^{-\frac{12}{s+7}}\text{ Hz},\quad &\text{for } k=0,\\
2.1\times10^{14}\varepsilon_{e,-1}^2\varepsilon_{B,-2}^{\frac{1}{2}}A_{10}^{\frac{s-1}{2(s+3)}}E_{\text{iso},53}^\frac{2}{s+3}\gamma_{\text{min},100}^\frac{2(s-1)}{s+3}\times\\\left(f_{s2,2}\right)^2f_{p,-1}^2(1+z)^\frac{2}{s+3}t_{h}^{-\frac{s+5}{s+3}}\text{ Hz},\quad &\text{for } k=2.
\end{cases}
\\
\nu^f_c &=\begin{cases}8.6\times10^{15}\varepsilon_{B,-2}^{-\frac{3}{2}}n_{-2}^{-\frac{3s+13}{2(s+7)}}E_{\text{iso},53}^{-\frac{4}{s+7}}\gamma_{\text{min},100}^\frac{4(1-s)}{s+7}\times\\\left(f_{s0,2}\right)^{-4}(1+z)^\frac{s-5}{s+7}t_h^{-\frac{2(s+1)}{s+7}}\text{ Hz},\quad &\text{for } k=0,\\
1.9\times10^{17
}\varepsilon_{B,-2}^{-\frac{3}{2}}A_{10}^{-\frac{3s+13}{2(s+3)}}E_{\text{iso},53}^{\frac{2}{s+3}}\gamma_{\text{min},100}^{\frac{2(s-1)}{s+3}}\times\\\left(f_{s2,2}\right)^{2}(1+z)^{-\frac{2s+4}{s+3}}t_{h}^\frac{s+1}{s+3}\text{ Hz},\quad &\text{for } k=2.
\end{cases}
\end{align}
\begin{align}
\label{eq:reverse4}
\nu^r_m &=\frac{(x-1)^4\sqrt{x^2+1}}{4x^2(x+1)(\gamma^f)^2}\nu_m^f\equiv\frac{f^\nu_{x}}{\left(\gamma^f\right)^2}\nu_m^f,\\&=\begin{cases}5.8\times10^{9}\varepsilon_{e,-1}^2\varepsilon_{B,-2}^{\frac{1}{2}}n_{-2}^{\frac{s+3}{2(s+7)}}E_{\text{iso},53}^\frac{2}{s+7}\gamma_{\text{min},100}^\frac{2(s-1)}{s+7}\times\\f^\nu_{x,-1}\left(f_{s0,2}\right)^2f_{p,-1}^2(1+z)^{-\frac{s+1}{s+7}}t_h^{-\frac{6}{s+7}}\text{ Hz},\quad &\text{for } k=0,\\
2.1\times10^{9}\varepsilon_{e,-1}^2\varepsilon_{B,-2}^{\frac{1}{2}}A_{10}^{\frac{1}{2}}f^\nu_{x,-1}f_{p,-1}^2t_h^{-1}\text{ Hz},\quad &\text{for } k=2.
\end{cases}
\\
\nu^r_c &= \frac{(x+1)^3}{(x^2+1)^\frac{3}{2}}\nu^f_c\overset{s=3}{\approx}2.5\nu^f_c,
\end{align}
where $t_h$ is the time measured in hours, $f^\nu_x\overset{s=3}{\approx}0.05[0.006]$ for $k=0[k=2]$, and
\begin{align}
    f_{s0}&=(1\text{hr})^{-\frac{3}{7+s}}\left[\frac{1.6\times10^{-24}x^{1+s}100^{1-s}}{(x^2-1)^2(s-1)}\right]^{-\frac{1}{7+s}}\overset{s=3}{\approx}52,\\
    f_{s2}&=(1\text{hr})^{-\frac{1}{3+s}}\left[\frac{2.7\times10^{-10}x^{1+s}100^{1-s}}{(x^2-1)^2(s-1)}\right]^{-\frac{1}{3+s}}\overset{s=3}{\approx}43,
\end{align}
In Appendix \ref{app:self}, we derive the self-absorption frequencies for both the FS and RS \citep[see also][]{nakar_early_2004}, resulting in
    \begin{align}
\nu_a^f
&=\begin{cases}1.6\times10^{9}\varepsilon_{e,-1}^{-1}\varepsilon_{B,-2}^{\frac{1}{5}}n_{-2}^{\frac{4s+20}{5(s+7)}}E_{\text{iso},53}^\frac{8}{5(s+7)}\gamma_{\text{min},100}^\frac{8(s-1)}{5(s+7)}\times\\f^f_{a,1}f_{s0,2}^\frac{8}{5}(1+z)^{-\frac{5s+11}{5(s+7)}}t_h^{\frac{3(s-1)}{5(s+7)}}\text{ Hz},\quad &\text{for } k=0,\\
6\times10^{8}\varepsilon_{e,-1}^{-1}\varepsilon_{B,-2}^{\frac{1}{5}}A_{10}^{\frac{4s+20}{5(s+3)}}E_{\text{iso},53}^{-\frac{8}{5(s+3)}}\gamma_{\text{min},100}^\frac{8(1-s)}{5(s+3)}\times\\f^f_{a,1}f_{s2,2}^{-\frac{8}{5}}(1+z)^{\frac{3s+1}{5(s+3)}}t_h^{-\frac{5s+7}{5(s+3)}}\text{ Hz},\quad &\text{for } k=2.
\end{cases}
\\
\nu^r_a &=\begin{cases}10^{11}\varepsilon_{e,-1}^\frac{2(p-1)}{p+4}\varepsilon_{B,-2}^{\frac{p+2}{2(p+4)}}n_{-2}^{\frac{(p+6)(s+3)}{2(p+4)(s+7)}}\times\\E_{\text{iso},53}^\frac{2(p+6)}{(p+4)(s+7)}\gamma_{\text{min},100}^\frac{2(p+6)(s-1)}{(p+4)(s+7)}f^r_{a0,11}f_{s0,2}^\frac{2(p+6)}{p+4}\times\\(1+z)^{\frac{6(p+6)}{(p+4)(s+7)}-1}t_h^{\frac{2(s+7)-6(p+6)}{(p+4)(s+7)}}\text{ Hz},\quad &\text{for } k=0,\\
10^{11}\varepsilon_{e,-1}^\frac{2(p-1)}{p+4}\varepsilon_{B,-2}^{\frac{p+2}{2(p+4)}}A_{10}^{\frac{p+6}{2(p+4)}}\times\\f^r_{a2,11}(1+z)^{\frac{4}{2(p+4)}}t_h^{-1}\text{ Hz},\quad &\text{for } k=2,
\end{cases}
\end{align}
where
\begin{align}
f_{a}^f&=\frac{(p+2)(p-1)f_p^{-\frac{5}{3}}}{3p+2}\overset{p=2.2}{\approx}9.7,\\
f^r_{a0}&\overset{p=2.2,s=3}{\approx}5.8\times10^{10},\\
f^r_{a2}&\overset{p=2.2,s=3}{\approx}8.7\times10^{10}.
\end{align}
For full expressions see Appendix \ref{app:self}.
The typical synchrotron specific power for a single electron is \citep{rybicki_radiative_1979}
\begin{equation}
    P_{\nu,\rm max}=P^\text{syn}_\nu(\nu_\text{syn})=0.7\frac{e^3 B\gamma^f}{m_ec^2},
\end{equation}
independent of the electron energy. The total number of radiating electrons in the swept-up external medium is (see Eq. (\ref{eq:Mism2}))
\begin{align}
        N^f_e&\approx\frac{M_\text{ext,iso}}{m_p}=\frac{4\pi A}{(3-k)m_p}\left(\frac{2\left(\gamma^f\right)^2ct}{1+z}\right)^{3-k},
        \end{align}
while for the shocked ejecta, it is given by (following Eq. (\ref{eq:Mism1}))
    \begin{align}
        \label{eq:reverse3}
    N^r_e&\approx\frac{M_\text{iso}(>\gamma_4)}{m_p}=\frac{4(3-k)x\gamma^f}{s(x^2-1)^2}N^f_e.
\end{align}
The peak observed flux is obtained by
\begin{equation}
    F_{\nu,\max}=\frac{(1+z)N_eP_{\nu,\rm max}}{4\pi D_L^2},
\end{equation}
where $D_L$ is the luminosity distance. For the FS, we have
\begin{equation}
\begin{aligned}
F_{\nu,\max}^f=\begin{cases}1.2\times10^{-1}\varepsilon_{B,-2}^{\frac{1}{2}}n_{-2}^\frac{3s+5}{2(s+7)}E^\frac{8}{s+7}_{\text{iso},53}\gamma_{\text{min},100}^\frac{8(s-1)}{s+7}D_{L,28}^{-2}\times\\
f_{s0,2}^{8}(1+z)^{\frac{10-2s}{s+7}}t_{h}^{\frac{3s-3}{s+7}}\text{ Jy},\quad &\text{for } k=0,\\
1.7\times10^{-2}\varepsilon_{B,-2}^{\frac{1}{2}}A_{10}^{\frac{3s+5}{2(s+3)}}E_{\text{iso},53}^\frac{2}{s+3}\gamma_{\text{min},100}^\frac{2(s-1)}{s+3}D_{L,28}^{-2}\times\\
f_{s2,2}^{2}(1+z)^\frac{s+5}{s+3}t_{h}^{-\frac{2}{s+3}}\text{ Jy},\quad &\text{for } k=2,
\end{cases}
\end{aligned}
\end{equation}
while for the RS
\begin{equation}
\begin{aligned}
F_{\nu,\max}^r&=\frac{4(3-k)x\sqrt{x^2+1}\gamma^f}{s(x^2-1)^{2}(x+1)}F_{\nu,\max}^f\equiv f_x^F\gamma^f F_{\nu,\max}^f,
\\&=\begin{cases}12\varepsilon_{B,-2}^{\frac{1}{2}}n_{-2}^\frac{3s+3}{2(s+7)}E^\frac{9}{s+7}_{\text{iso},53}\gamma_{\text{min},100}^\frac{9(s-1)}{s+7}D_{L,28}^{-2}\times\\
f_x^Ff_{s0,2}^{9}(1+z)^{\frac{13-2s}{s+7}}t_{h}^{\frac{3s-6}{s+7}}\text{ Jy},\quad &\text{for } k=0,\\
1.7\varepsilon_{B,-2}^{\frac{1}{2}}A_{10}^{\frac{3s+3}{2(s+3)}}E_{\text{iso},53}^\frac{3}{s+3}\gamma_{\text{min},100}^\frac{3(s-1)}{s+3}D_{L,28}^{-2}\times\\
f_x^Ff_{s2,2}^{3}(1+z)^\frac{s+6}{s+3}t_{h}^{-\frac{3}{s+3}}\text{ Jy},\quad &\text{for } k=2,
\end{cases}
\end{aligned}
\end{equation}
where $f_x^F\overset{s=3}{\approx}0.6[0.85]$ for $k=0[k=2]$. Using the expressions derived above, the observed synchrotron flux can be computed following the standard prescription \citep{sari_spectra_1998}. For the FS: $\nu^f_a<\nu^f_m<\nu^f_c$:
\begin{equation}
    F^f_\nu=\begin{cases}

        \left(\frac{\nu}{\nu^f_m}\right)^\frac{1}{3}F^f_{\nu,\max},\quad &\text{for } \nu<\nu^f_m,\\
        \left(\frac{\nu}{\nu^f_m}\right)^\frac{1-p}{2}F^f_{\nu,\max},\quad &\text{for } \nu^f_m<\nu<\nu^f_c,\\
        \left(\frac{\nu^f_c}{\nu^f_m}\right)^\frac{1-p}{2}\left(\frac{\nu}{\nu^f_c}\right)^{-\frac{p}{2}}F^f_{\nu,\max},\quad &\text{for } \nu^f_c<\nu,
    \end{cases}
\end{equation}
while for the RS: $\nu^r_m<\nu^r_a<\nu^r_c$:\begin{equation}
    F^r_\nu=\begin{cases}
        \left(\frac{\nu^r_a}{\nu^r_m}\right)^\frac{1-p}{2}\left(\frac{\nu}{\nu^r_a}\right)^\frac{5}{2}F^r_{\nu,\max},\quad &\text{for } \nu<\nu^r_a,\\
        \left(\frac{\nu}{\nu^r_m}\right)^{\frac{1-p}{2}}F^r_{\nu,\max},\quad &\text{for } \nu^r_a<\nu<\nu^r_c,\\
        \left(\frac{\nu^r_c}{\nu^r_m}\right)^\frac{1-p}{2}\left(\frac{\nu}{\nu^r_c}\right)^{-\frac{p}{2}}F^r_{\nu,\max},\quad &\text{for } \nu^r_c<\nu.
    \end{cases}
\end{equation}
The temporal and spectral evolution of the observed flux depends on both the external density profile, parametrized by $k$, and the relevant spectral regime. Adopting the standard parametrization $F_\nu \propto t^{-\alpha}\nu^{-\beta} $, we summarize the corresponding closure relations between the temporal decay index $\alpha$ and the spectral index $\beta$ in the cases of a uniform ISM ($k=0$), a wind profile produced by a constant mass-loss rate ($k=2$), and for an arbitrary $k$, for both the FS and RS in Table~\ref{tab:closure} \citep[see also][]{sari_impulsive_2000}.
\begin{table*}
\caption{Temporal and spectral indices of the synchrotron flux,
$F_\nu \propto t^{-\alpha}\nu^{-\beta} $, expected from both the FS and RS emission resulting from the hydrodynamic evolution derived in \S~\ref{sec:hydro}. $s\rightarrow\infty$ corresponds to the coasting phase ($\gamma\sim$const), while $s\rightarrow 1$ corresponds to the adiabatic BM phase, where most of the energy is in the shocked external medium.}
\label{tab:closure}
\begin{tabular}{cccccc}
\hline
Emission region & Spectral regime & $\beta$ & $\alpha\ (k=0)$ & $\alpha\ (k=2)$ & $\alpha(k)$\\
\hline \multirow{6}{*}{Forward shock}
& $\nu < \nu_m$
& $-\dfrac{1}{3}$
& $-\dfrac{3s+1}{s+7}$
& $\dfrac{1-s}{3(s+3)}$ & $\dfrac{4ks+2k-9s-3}{3(s+7-2k)}$\\[6pt]
& $\nu_m < \nu < \nu_c$
& $\dfrac{p-1}{2}$
& $\dfrac{6p-3s-3}{s+7}$
& $\dfrac{5p+sp-s-1}{2(s+3)}$ & $\dfrac{kps-7kp+5ks+5k+24p-12s-12}{4(s+7-2k)}$\\[6pt]
& $\nu > \nu_c$
& $\dfrac{p}{2}$
& $\dfrac{6p-2s-2}{s+7}$
& $\dfrac{5p+sp-2s-2}{2(s+3)}$ & $\dfrac{kps-7kp+2ks+2k+24p-8s-8}{4(s+7-2k)}$\\
\hline \multirow{6}{*}{Reverse shock}
& $\nu < \nu_a$
& $-\dfrac{5}{2}$
& $-\dfrac{2s+8}{s+7}$
& $-\dfrac{5s+9}{2(s+3)}$ & $\dfrac{-ks+7k-8s-32}{4(s+7-2k)}$\\[6pt]
& $\nu_a < \nu < \nu_c$
& $\dfrac{p-1}{2}$
& $\dfrac{3p-3s+3}{s+7}$
& $\dfrac{3p+sp-s+3}{2(s+3)}$ & $\dfrac{kps-3kp+5ks-3k+12p-12s+12}{4(s+7-2k)}$\\[6pt]
& $\nu > \nu_c$
& $\dfrac{p}{2}$
& $\dfrac{3p-2s+4}{s+7}$
& $\dfrac{3p+sp-2s+2}{2(s+3)}$ & $\dfrac{kps-3kp+2ks-6k+12p-8s+16}{4(s+7-2k)}$\\
\hline
\end{tabular}
\end{table*}
After the crossing time, the FS emission is expected to follow the standard decelerating blast wave temporal properties, which are given in Table \ref{tab:closure}, by taking $s=1$ \citep{sari_spectra_1998,zhang_gamma-ray_2004,shen_coasting_2011}.
\section{Constraints from the observed properties of X-ray plateaus}
\label{sec:obs}
In this section, we demonstrate that the stratified-outflow model developed in $\S\ref{sec:hydro}$-\ref{sec:emission} reproduces the typical observed properties of X-ray plateaus, while providing a prediction for the emergent emission of the long-lived RS. Rather than modeling individual bursts, we translate the phenomenology of plateau samples into generic observational constraints and examine whether these are satisfied for typical model parameters.
X-ray plateaus observed by Swift exhibit several robust and well-established properties \citep[e.g.,][]{nousek_evidence_2006,zhang_physical_2006,shen_coasting_2011,ronchini_combined_2023}. During the plateau phase, the X-ray flux decays slowly, $F_X \propto t^{-\alpha_{X,1}}$, $\alpha_{X,1} \sim 0.3$--$0.6$, with typical observed flux of $F_\nu(1\,\mathrm{keV}) \sim 0.1$--$10~\mu\mathrm{Jy}.$ The plateau ends at a break time $t_b \sim 10^{3}$--$10^{5}\,\mathrm{s}$, and is followed by a smooth transition to a steeper decay with a temporal index $\alpha_{X,2} \sim 1.0$--$1.5,$ consistent with standard external-shock evolution.
Spectrally, the X-ray emission during the plateau is well described by a single power law, $F_\nu \propto \nu^{-\beta_{X,1}}$, with a typical spectral index $\beta_{X,1} \sim 1.0$--$1.2$, corresponding to an electron power-law index of $p \sim 2.0$--$2.4$. Importantly, the spectral index shows little or no evolution across the plateau break, $\beta_{X,1} \simeq \beta_{X,2},$ in the majority of events, indicating that the break is dynamical rather than spectral in origin.
Any explanation must therefore account for the above-mentioned properties simultaneously.
\subsection{Plateau slope}
In the stratified-outflow model, the ejecta mass above LF $\gamma$ follows a power-law distribution $M(>\gamma)\propto\gamma^{-s}$. As slower ejecta catches up with the decelerating blast wave, the total energy of the shocked region increases with time, modifying the synchrotron temporal decay. Within the model framework, the X-ray emission is dominated by the FS emission. For a given external density profile and spectral regime, the temporal index $\alpha_X$ can be written explicitly as a function of $s$ and the electron power-law index $p$ (see Table \ref{tab:closure}). We focus on the case in which the X-ray band lies above both the peak and cooling frequencies,
$\nu_X > \max(\nu_m,\nu_c)$, as motivated by the observed X-ray spectral slopes $\beta_X \simeq 1.0$--$1.2$. For representative values $p\simeq2$--$2.4$ and ISM-like external medium ($k = 0$), stratification indices of
\begin{equation}
s \simeq 2.5\text{--}4,
\end{equation}
naturally yield $\alpha_X \simeq 0.3$--$0.6$, consistent with the observed plateau distribution. For a wind-like external medium ($k=2$), reproducing the same observed plateau slopes requires larger inferred values, $s > 4$. We stress, however, that the ejecta stratification index, $s$, is an intrinsic property of the outflow and does not itself depend on the external medium. Rather, because the closure relations depend on both $s$ and $k$ (Table \ref{tab:closure}), the value of $s$ inferred from a given observed plateau slope depends on the assumed external density profile. In particular, for the same ejecta stratification and electron index, $p$, a wind-like external medium produces a steeper decline than an ISM-like medium. Accordingly, the wind and ISM cases should not be interpreted as implying different intrinsic ejecta populations, but rather as different mappings between the observed plateau properties and the underlying stratification index.
\subsection{Plateau duration}
The plateau duration corresponds to the time over which slower ejecta continues to contribute energy to the shock. Once this reservoir is exhausted, the system approaches the standard self-similar deceleration regime, producing a smooth transition to the post-plateau decay without an abrupt flux drop. The plateau duration is therefore
set by the RS crossing time $t_{\rm cross}$ (Eq.~\eqref{eq:t}).
\begin{itemize}
    \item ISM ($k=0$) case:

For a given stratification index $s\simeq 2.5$--$4.0$, inferred from the plateau slopes, we find $h_\gamma\simeq 3$--$10$. The observed plateau duration $t_b\sim 10^3$--$10^5$ s leads to
\begin{equation}
    (1+z)E^\frac{1}{3}_{\text{iso},53}n^{-\frac{1}{3}}_{-2}\gamma^{-\frac{8}{3}}_{\text{min},100}\simeq \frac{t_b}{9\times10^2h_\gamma(s)}\lesssim1.\end{equation}
For typical parameters of long GRBs, $z\simeq1.5$--$2.5,E_{\text{iso},53}\approx1,n_{-2}\simeq10$--$100$, we find
\begin{equation}
\gamma_{\min}\sim 70-100.
\end{equation}
The Lorentz factor of the forward-shocked plasma at the end of the plateau is smaller than the minimum ejecta Lorentz factor by a factor $x$, namely $\gamma^f(t_b)=\gamma_\text{min}/x$, and is therefore typically only a few tens, consistent with the standard BM estimate at $t_b$.
\item Wind ($k=2$) case:

For $s>4$, inferred from the plateau slopes, we find $h_a\simeq0.3$--$0.5$. The observed plateau duration $t_b\sim 10^3$--$10^5$ s leads to
\begin{equation}
(1+z)E_{\text{iso},53}A^{-1}_{10}\gamma^{-4}_{\text{min},100}\simeq \frac{t_b}{3.3\times10^2h_a(s)}\lesssim75.
\end{equation}
For typical parameters of $z\simeq1.5$--$2.5,E_{\text{iso},53}\approx1,A_{10}\simeq1$, a lower value of $\gamma_{\min}= x\gamma^f(t_b)\sim 40$ is required to match the observed plateau duration, consistent with the coasting-in-wind model \citep{shen_coasting_2011}. The combination of $s>4$ and $\gamma_\text{min}\sim40$ implies that most of the ejecta energy resides in the slowest material, while the fraction carried by ejecta with LF $>100$ is strongly suppressed.
In the remainder of this work, we focus on the ISM case, which is independently supported by the plateau flux and spectral constraints discussed below.
\end{itemize}
\subsection{Spectral regime}
We consider the regime in which the observed X-ray band lies above both the peak and cooling frequencies, $\nu_X > \max(\nu_m,\nu_c)$, as motivated by the observed X-ray spectral slopes $\beta_X\simeq1.0$--$1.2$. Using the expressions for $\nu^f_m$ and $\nu^f_c$ derived in \S~\ref{sec:emission}, we now examine the conditions under which this ordering is satisfied during
the plateau phase.
For stratification indices
$s\simeq2.5$--$4.0$ and observation time of $t\sim10^3$~s, the expression for $\nu^f_m$ shows that even for maximal choice of parameters (e.g., $\varepsilon_e=0.3$, $\varepsilon_B=10^{-2}$), $\nu^f_m$ lies well below the X-ray band already at the beginning of the plateau. As a result, the condition $\nu^f_m<\nu_X$ is naturally satisfied over the relevant parameter space and does not impose an additional constraint.
The cooling frequency provides a more restrictive bound. Using the expression for $\nu_c$, the requirement that the X-ray band remains above the cooling break during the plateau imposes a lower bound on the magnetic energy fraction $\varepsilon_B$. Using the inferred values of $s\simeq2.5$--$4.0$, $\gamma_{\min}\simeq100$, and typical parameters of $z\simeq1.5$--$2.5,E_{\text{iso},53}\approx1,n_{-2}\simeq10$--$100$, this condition implies
\begin{equation}
\varepsilon_B \gtrsim 10^{-3}.
\end{equation}
Both characteristic frequencies decline monotonically with time during the plateau phase. Consequently, if
the condition $\nu_X>\max(\nu_m,\nu_c)$ is satisfied at $t\sim10^3$~s, it remains satisfied throughout the
plateau.
\subsection{Plateau flux}
In the regime  $\nu_X>\max(\nu_m,\nu_c)$, the FS synchrotron emission is
\begin{equation}
\label{eq:x}
F_{X} =
F^f_{\nu,\max}\,
\left(\nu^f_c\right)^{1/2}\,
\left(\nu_m^f\right)^{\frac{p-1}{2}}\,
\nu_X^{-p/2},
\end{equation}
which explicitly states that the X-ray flux is determined by the combination of $F^f_{\nu,\max}$, $\nu^f_m$, and $\nu^f_c$.
Substituting the expressions for $F_{\nu,\max}$, $\nu_m$, and $\nu_c$ derived in \S~\ref{sec:emission}, and fixing
$\gamma_{\min}=100$, $s\simeq2.5$--$4.0$ shows that for $p\simeq2.0$--$2.4$ the X-ray flux depends primarily on the electron energy fraction $\varepsilon_e$ and the isotropic-equivalent energy $E_{\rm iso}$. The dependence on $\varepsilon_B$ and the external density normalization is weaker due to compensating scalings between $F^f_{\nu,\max}$, $\nu^f_m$, and $\nu^f_c$,
while the dominant redshift dependence enters through the luminosity distance $D_L^{-2}(z)$.
Observed X-ray plateaus have fluxes $F_{\nu,X}(1\,\mathrm{keV})\sim0.1$--$10\,\mu\mathrm{Jy}$ at observer
times $t\sim10^3$--$10^5$~s. Matching this range for $z\simeq1.5$--$2.5$ yields
\begin{equation}
\varepsilon_e \sim 0.05\text{--}0.5,
\end{equation}
for typical GRB energetics, with $E_{\rm iso} \sim 10^{52.5}\text{--}10^{53.5}\,\mathrm{erg}$. Lower values of $\varepsilon_e$ underproduce the observed X-ray flux, while
significantly larger values require unusually low energies. Additionally, joint X-ray and optical analyses show that, for the majority of events, the temporal and spectral properties are consistent with a common synchrotron origin in a refreshed FS \citep{li_statistical_2026}.
The stratified-outflow model presented here is also consistent with the Dainotti relation \citep{dainotti_timeluminosity_2008,dainotti_discovery_2010} between the X-ray plateau luminosity and its end time. In our framework, the plateau duration (identified with the reverse-shock crossing time) is most sensitive to the minimum LF of the ejecta, $\gamma_\text{min}$. For example, combining Eq. (\ref{eq:t}) and Eq. (\ref{eq:x}) for typical values of $s=3,p=2$ yields
\begin{equation}
F_X\propto\varepsilon_eE_\text{iso}(1+z)t_\text{cross}^{-1}D_L^{-2}.
\end{equation}
This anti-correlation between the X-ray luminosity at the end of the plateau and its duration is consistent with the observations. Given the additional variations in electron power index and the detailed ejecta stratification, the relation is naturally expected to exhibit a significant scatter.
\subsection{Post-plateau decline}
In the majority of GRBs, the decay following the plateau phase is broadly consistent with the standard adiabatic evolution of an external FS, with post-plateau temporal slopes falling within the range expected for normal afterglow decay. This behavior is observed in both X-ray and optical bands when the emission is consistent with a common external-shock origin, reinforcing the interpretation that many plateaus are associated with refreshed shocks rather than a fundamentally distinct emission mechanism.
A smaller subset of bursts, however, exhibits a post-plateau decline, observed by Swift-XRT, that is significantly steeper than predicted by simple FS models \citep[e.g., GRB 060614A, GRB 050730A, GRB 120404A, GRB 130408A, GRB 130609B, GRB 170317A, GRB
180224A, and GRB 180620A,][]{swain_grb_2025,ror_investigating_2025}. In these cases, the steepening may be consistent with a jet break near the end of the plateau phase, so that the light curve transitions directly from the plateau into the jet-spreading regime. This interpretation is particularly plausible for plateaus associated with lower effective LFs of the forward-shocked plasma at the end of the plateau, $\gamma^f=\gamma_\text{min}/x$, increasing the probability of an early jet break. Such a scenario naturally links the plateau properties to the jet geometry and does not require invoking an internal origin for the emission.
\subsection{Reverse shock emission}
\begin{figure}
\includegraphics[width=\columnwidth]{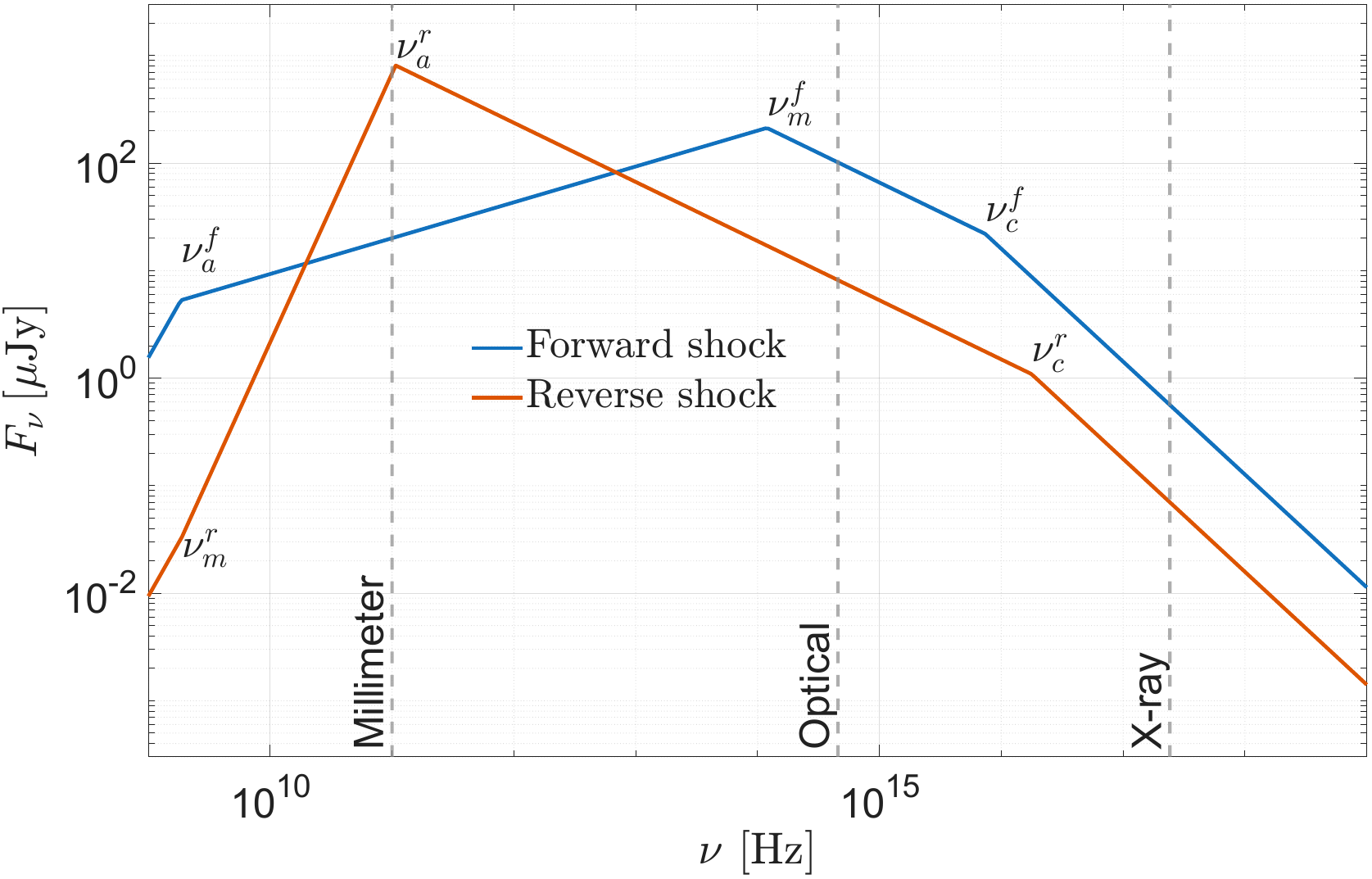}
    \caption{Representative broadband synchrotron spectra of the FS (blue) and RS (red) at $t\simeq1$ hr for fiducial parameters $p=2.1$, $\varepsilon_e=0.1$, $\varepsilon_B=10^{-2}$, $n=10^{-1}\,\mathrm{cm^{-3}}$, $E=10^{53}$ erg, $\gamma_{\min}=100$, $s=3$, and $z=2$. Vertical dashed lines indicate mm, optical, and X-ray bands. The characteristic frequencies of the FS and RS are also provided. The FS dominates the optical and X-ray emission, whereas the RS peaks at mm wavelengths and outshines the FS at low frequencies, producing a bright, long-lived mm component during the X-ray plateau phase.}
    \label{fig:spectrum}
\end{figure}

\begin{figure}
\includegraphics[width=\columnwidth]{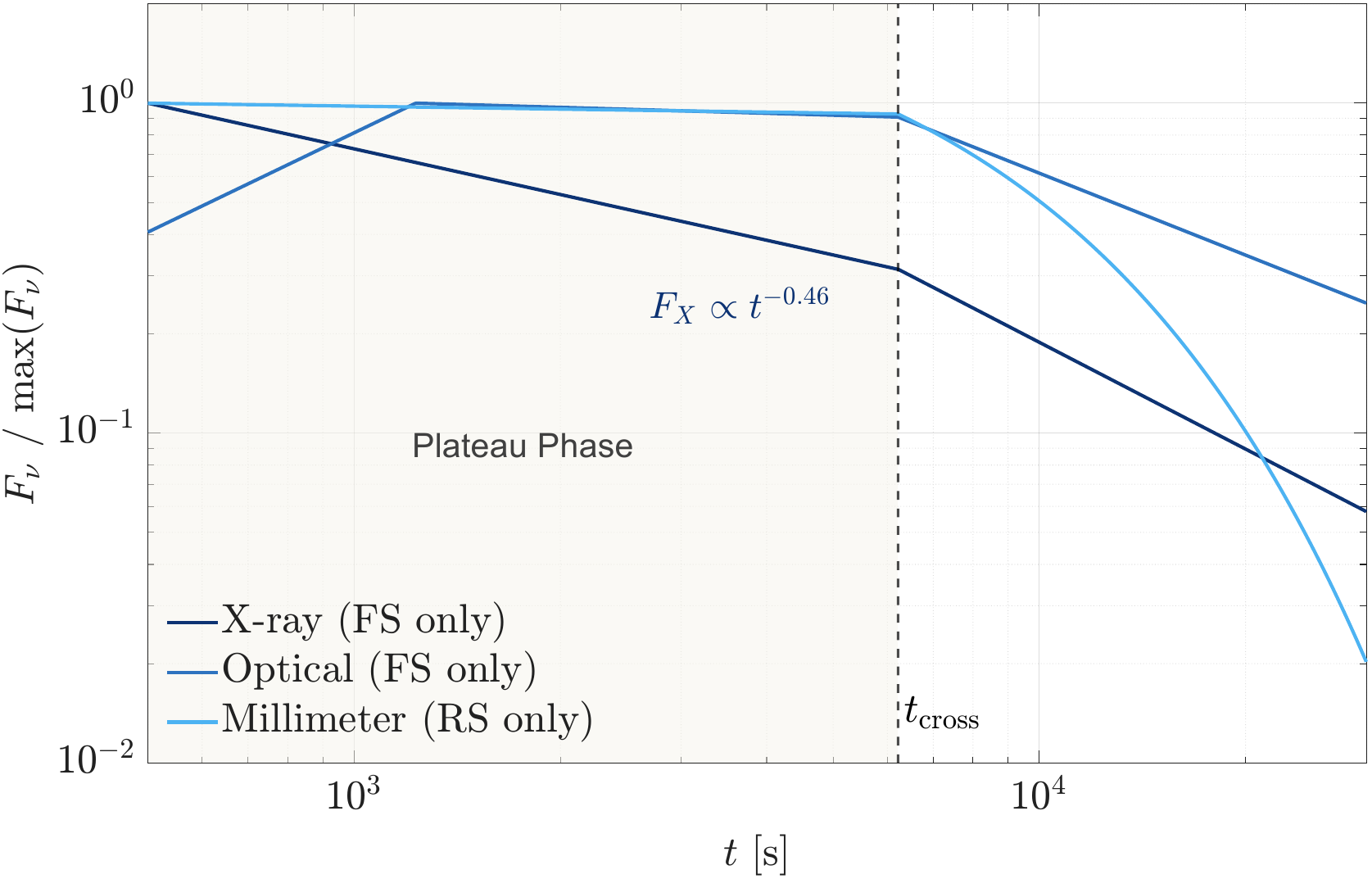}
    \caption{Representative synchrotron X-ray, optical and mm light curves for fiducial parameters $p=2.1$, $\varepsilon_e=0.1$, $\varepsilon_B=10^{-2}$, $n=10^{-1}\,\mathrm{cm^{-3}}$, $E=10^{53}$ erg, $\gamma_{\min}=100$, $s=3$, and $z=2$. As shown in Fig. \ref{fig:spectrum}, the FS dominates the optical and X-ray emission, whereas the RS peaks at mm wavelengths and outshines the FS at low frequencies, producing a bright, long-lived mm component during the X-ray plateau phase. After the reverse shock crossing, the X-ray and optical light curves follow the typical decline from FS with standard adiabatic evolution, whereas the RS emission should decline exponentially.}
    \label{fig:lc}
\end{figure}

The parameter ranges inferred from the X-ray plateau analysis have direct and robust implications for the RS emission. While the FS naturally dominates the X-ray band, the RS generically produces a much brighter component at low frequencies, with its peak emission falling in the mm band during the plateau phase.
This behavior follows from two general properties of the RS in stratified outflows. First, at a given observer time, the RS processes a substantially larger number of electrons than the FS, since it sweeps through the ejecta rather than only the external material (Eq. \eqref{eq:reverse3}). Second, the typical electron LF in the shocked ejecta is significantly lower than in the FS (Eq. (\ref{eq:Lor})), which shifts the characteristic synchrotron frequencies of the RS to much lower values (Eq. (\ref{eq:reverse4})). As a result, the RS spectrum peaks at frequencies well below those of the FS, while maintaining a comparable or larger peak flux \citep{sari_impulsive_2000}.
Fig.~\ref{fig:spectrum} illustrates a representative broadband spectrum of the FS and RS at $t\simeq1$ hr, while Fig. \ref{fig:lc} illustrates representative light curves in the X-ray, optical, and mm bands, during (and shortly after) the plateau phase. Both are shown for fiducial parameters consistent with our inferred bounds: $p=2.1$, $\varepsilon_e=0.1$, $\varepsilon_B=10^{-2}$, $n=10^{-1}\,\mathrm{cm^{-3}}$, $E=10^{53}$ erg, $\gamma_{\min}=100$, $s=3$, and $z=2$. In these examples, the FS emission peaks in the optical band and dominates the X-ray emission, while the RS peak lies in the mm range, where it outshines the FS by more than an order of magnitude. At higher frequencies, the RS contribution declines rapidly and becomes negligible in the X-ray band. The prominence of the RS in the mm band is therefore a robust consequence of the ejecta stratification required to reproduce the X-ray plateau slopes. During the plateau phase, the RS mm emission is expected to persist as long as the RS crosses progressively slower ejecta, and to fade once this process ceases. This behavior implies a clear, testable prediction: GRBs exhibiting X-ray plateaus should be accompanied by a bright, long-lived mm component that is absent or significantly weaker once the plateau ends. Thus, early ($\sim$ hour) and sustained mm observations, particularly with facilities such as AtLAST \citep{mroczkowski_conceptual_2025} or ALMA \citep{laskar_first_2018,laskar_alma_2019}, can provide a sensitive probe of the stratified-ejecta interpretation of X-ray plateaus, complementary to constraints obtained from the X-ray band alone.
\section{Conclusions}
\label{sec:conclusions}
In this work, we developed a novel analytic framework for treating the hydrodynamics of UR, radially stratified outflows interacting with an external medium. By explicitly accounting for a continuous distribution of LFs within the ejecta, we derived closed-form expressions describing the dynamics of a long-lived, mildly relativistic RS propagating through the outflow (\S \ref{sec:hydro}). Within this framework, we inferred the RS crossing time (Eq. (\ref{eq:t})), which is $\sim$ an order of magnitude larger than previous estimates based on the thin-shell approximation for an ISM-like environment. Then, in \S \ref{sec:emission}, we computed the resulting FS and RS synchrotron emission, focusing in particular on a constant-density ISM ($k=0$) and a wind-like ($k=2$) external medium.
We find that the interaction of stratified relativistic ejecta with the external medium naturally produces FS emission with a long-lasting, shallow X-ray decay, whose duration and temporal slope are consistent with the observed properties of X-ray plateaus, without invoking an additional energy reservoir beyond the initial ejecta, a wind-like external medium, or an additional high-energy emission component (\S~\ref{sec:obs}). The appearance of a plateau is associated with an intermediate range of the stratification index $s$ \citep[similar to][]{granot_distribution_2006}. In practice, the cumulative ejecta distribution $M(>\gamma)$ (or equivalently $E(>\gamma)$) is not expected to follow a single power law over the entire LF range: by construction, it must be steep at the largest LFs and gradually flatten toward the smallest LFs. In our model, an effective slope in the range $2.5 \lesssim s \lesssim 4$ during the relevant phase leads to a prolonged RS crossing time and, consequently, a shallow X-ray decay. For steeper stratification ($s \gg 4$), the crossing time approaches the thin-shell limit, producing a more impulsive evolution with a rapid transition to the standard adiabatic blast wave decay.
The model is also consistent with the observed anti-correlation between X-ray plateau luminosity and duration \citep{dainotti_timeluminosity_2008,dainotti_discovery_2010}, without requiring additional assumptions about prolonged central-engine activity. The RS associated with the same stratified ejecta is predicted to generate a bright, long-lived emission component at lower frequencies. For the parameter ranges inferred from the plateau slopes and durations, the RS spectrum peaks in the mm band and outshines the FS emission at these wavelengths during the plateau phase, while remaining negligible in the X-ray band. As a result, mm observations during the plateau phase can provide a particularly stringent test of the stratified-ejecta scenario, since the FS alone is expected to be faint at these wavelengths \citep[e.g.,][]{laskar_first_2018}. Conversely, the absence of a mm excess during the plateau would place strong constraints on the allowed microphysical parameters or on the role of ejecta stratification in
shaping the observed X-ray plateaus.
The transition from the plateau phase to the standard adiabatic blast wave evolution occurs smoothly once the RS  has crossed the slowest ejecta, and the resulting post-plateau temporal and spectral behavior is consistent with that inferred from broadband afterglow observations. However, a small fraction of events shows a significantly steeper post-plateau decline, which in some cases may be explained by a jet break occurring close to the end of the plateau \citep{ror_investigating_2025}. Taken together, the plateau flux level shows that once the ejecta stratification index $s$ and the minimum LF $\gamma_{\min}$ are fixed by the plateau slope and duration, the resulting microphysical parameters and energetics fall within standard ranges (\S~\ref{sec:obs}). Additionally, the same region of parameter space that reproduces the plateau morphology also satisfies the required spectral ordering throughout the plateau phase, without the need for fine-tuning.
A key aspect of this model is that the stratified structure required to power the plateau is expected on physical grounds. UR outflows (with LF $\gtrsim100$) responsible for the prompt $\gamma$-ray emission are unlikely to be launched with a single LF. Instead, variability in the central engine and internal dissipation naturally lead to ejecta with a broad LF distribution. In addition, modestly off-axis viewing angles could further flatten and extend the observed light curve, enhancing the plateau without qualitatively altering the underlying dynamics \citep{beniamini_afterglow_2020}. Taken together, this model provides a unified explanation in which the same outflow that produces the prompt emission also gives rise to the X-ray plateau and its subsequent transition to the BM phase. The plateau thus reflects the hydrodynamic evolution of a stratified relativistic outflow rather than prolonged central-engine activity or fine-tuned geometric effects, offering a physically economical interpretation of one of the most persistent features of GRB afterglows.
An additional strength of this framework is that it provides a consistent energy budget for the plateau phase without introducing tensions in the inferred radiative efficiency. Because the energy that powers the plateau is already carried by the same UR outflow responsible for the prompt emission, there is no need to invoke a large reservoir of previously hidden kinetic energy or sustained late-time energy injection. The gradual release of energy from the stratified ejecta naturally produces the plateau while maintaining prompt radiative efficiencies within the range inferred from observations.
\textit{Note added during review.} After submission of this manuscript, \citet{keating_rapid-response_2026} reported rapid-response 1.3 mm observations of GRB 260127A, detecting a bright mm counterpart within $\sim13$ minutes of the burst that declined rapidly on day timescales. This behavior is broadly consistent with the early mm component predicted in this work.
\section*{Acknowledgments}
We thank Eli Waxman and Jonathan Granot for helpful discussions. Additionally, we thank the anonymous referee for useful comments, which improved the quality of the manuscript. We also thank the Yukawa Institute for Theoretical Physics at Kyoto University. Discussions during the Yukawa International Seminar YKIS2026a on "Black Holes and Neutron Stars with Multi-Messengers" were useful to complete this work. This work was in part supported by Grant-in-Aid for Scientific Research (grant No.~23H01169, 23H04900) of Japanese MEXT/JSPS and the JST FOREST Program (JPMJFR2136).
\section*{Data Availability}
The data underlying this article will be shared following a reasonable request to the corresponding author.
\bibliographystyle{mnras}
\bibliography{references}

@misc{keating_rapid-response_2026,
	title = {Rapid-response 1.3 mm {Observations} of {GRB} {260127A} with the {Submillimeter} {Array}},
	url = {http://arxiv.org/abs/2604.14297},
	doi = {10.48550/arXiv.2604.14297},
	urldate = {2026-04-23},
	publisher = {arXiv},
	author = {Keating, Garrett K. and Laskar, Tanmoy and Ho, Anna Y. Q. and Blanchard, Peter K. and Alexander, Kate D. and Berger, Edo and Gurwell, Mark and Eftekhari, Tarraneh and Xu, Chloe T. and Lovell, Joshua Bennett and Rao, Ramprasad and Williams, Peter K. G.},
	month = apr,
	year = {2026},
	note = {arXiv:2604.14297 [astro-ph]},
}

@article{laskar_alma_2019,
	title = {{ALMA} {Detection} of a {Linearly} {Polarized} {Reverse} {Shock} in {GRB} {190114C}},
	volume = {878},
	issn = {0004-637X},
	url = {https://ui.adsabs.harvard.edu/abs/2019ApJ...878L..26L},
	doi = {10.3847/2041-8213/ab2247},
	urldate = {2026-03-03},
	journal = {The Astrophysical Journal},
	publisher = {IOP},
	author = {Laskar, Tanmoy and Alexander, Kate D. and Gill, Ramandeep and Granot, Jonathan and Berger, Edo and Mundell, C. G. and Barniol Duran, Rodolfo and Bolmer, J. and Duffell, Paul and van Eerten, Hendrik and Fong, Wen-fai and Kobayashi, Shiho and Margutti, Raffaella and Schady, Patricia},
	month = jun,
	year = {2019},
	note = {ADS Bibcode: 2019ApJ...878L..26L},
	pages = {L26},
}

@article{yamazaki_prior_2009,
	title = {Prior {Emission} {Model} for {X}-ray {Plateau} {Phase} of {Gamma}-{Ray} {Burst} {Afterglows}},
	volume = {690},
	issn = {0004-637X},
	url = {https://ui.adsabs.harvard.edu/abs/2009ApJ...690L.118Y},
	doi = {10.1088/0004-637X/690/2/L118},
	urldate = {2026-03-03},
	journal = {The Astrophysical Journal},
	publisher = {IOP},
	author = {Yamazaki, Ryo},
	month = jan,
	year = {2009},
	note = {ADS Bibcode: 2009ApJ...690L.118Y},
	pages = {L118--L121},
}

@article{laskar_first_2018,
	title = {First {ALMA} {Light} {Curve} {Constrains} {Refreshed} {Reverse} {Shocks} and {Jet} {Magnetization} in {GRB} {161219B}},
	volume = {862},
	issn = {0004-637X},
	url = {https://ui.adsabs.harvard.edu/abs/2018ApJ...862...94L},
	doi = {10.3847/1538-4357/aacbcc},
	urldate = {2026-03-03},
	journal = {The Astrophysical Journal},
	publisher = {IOP},
	author = {Laskar, Tanmoy and Alexander, Kate D. and Berger, Edo and Guidorzi, Cristiano and Margutti, Raffaella and Fong, Wen-fai and Kilpatrick, Charles D. and Milne, Peter and Drout, Maria R. and Mundell, C. G. and Kobayashi, Shiho and Lunnan, Ragnhild and Barniol Duran, Rodolfo and Menten, Karl M. and Ioka, Kunihito and Williams, Peter K. G.},
	month = aug,
	year = {2018},
	note = {ADS Bibcode: 2018ApJ...862...94L},
	pages = {94},
}

@article{laskar_energy_2015,
	title = {Energy {Injection} in {Gamma}-{Ray} {Burst} {Afterglows}},
	volume = {814},
	issn = {0004-637X},
	url = {https://ui.adsabs.harvard.edu/abs/2015ApJ...814....1L},
	doi = {10.1088/0004-637X/814/1/1},
	urldate = {2026-03-03},
	journal = {The Astrophysical Journal},
	publisher = {IOP},
	author = {Laskar, Tanmoy and Berger, Edo and Margutti, Raffaella and Perley, Daniel and Zauderer, B. Ashley and Sari, Re'em and Fong, Wen-fai},
	month = nov,
	year = {2015},
	note = {ADS Bibcode: 2015ApJ...814....1L},
	pages = {1},
}

@misc{li_minutes-long_2026,
	title = {Minutes-long soft {X}-ray prompt emission from a compact object merger},
	url = {https://ui.adsabs.harvard.edu/abs/2026arXiv260114137L},
	doi = {10.48550/arXiv.2601.14137},
	urldate = {2026-02-27},
	publisher = {arXiv},
	author = {Li, An and Wang, Chen-Wei and Passaleva, Niccolò and An, Jie and Zhang, Bin-Bin and Troja, Eleonora and Yin, Yi-Han Iris and Liu, Yuan and Xiong, Shao-Lin and Xin, Li-Ping and Shao, Yi-Xuan and Yang, Jun and Sun, Hui and Xu, Dong and Yang, Yu-Han and Ricci, Roberto and Gao, He and Antier, Sarah and Becerra, Rosa L. and Cao, Jia-Xin and Castro-Tirado, Alberto Javier and Chen, Xin-Lei and Cheng, Ye-Hao and Chen, Yong and Cheng, Hua-Qing and D'Elia, Valerio and De Pasquale, Massimiliano and Dong, Yong-Wei and Elhosseiny, Eslam and Eyles-Ferris, Rob A. J. and Gritsevich, Maria and Han, Xu-Hui and Hartmann, Dieter and Hu, You-Dong and Hu, Jing-Wei and Jia, Shu-Mei and Kochiashvili, Nino and Lei, Wei-Hua and Levan, Andrew J. and Li, Cheng-Kui and Li, Dong-Yue and Li, Hua-Li and Li, Xiao-Bo and Ling, Zhi-Xing and Liu, He-Yang and Lv, Hou-Jun and Malesani, Daniele B. and O'Connor, Brendan and Pan, Hai-Wu and Bhushan Pandey, Shashi and Perez-Garcia, Ignacio and Pieterse, Daniëlle L. A. and Pillas, Marion and Qiu, Yu-Lei and Saccardi, Andrea and Sánchez-Ramírez, Rubén and Tan, Wen-Jun and Tanasan, Manasanun and Tanvir, Nial R. and Vergani, Susanna D. and Wang, Jing and Wang, Xiao-Feng and Wu, Qin-Yu and Yi, Shu-Xu and Yusufjon, Tillayev and Zhang, Chen and Zhang, Wen-Da and Zhang, Yi-Jia and Zhao, Guo-Ying and Zheng, Chao and Zheng, Shi-Jie and Zhou, Chang and Zhou, Ping and Cordier, Bertrand and Wei, Jian-Yan and Yuan, Weimin and Zhang, Shuang-Nan and Zhang, Bing},
	month = jan,
	year = {2026},
	note = {ADS Bibcode: 2026arXiv260114137L},
}

@article{kisaka_long-lasting_2015,
	title = {Long-lasting {Black} {Hole} {Jets} in {Short} {Gamma}-{Ray} {Bursts}},
	volume = {804},
	issn = {0004-637X},
	url = {https://ui.adsabs.harvard.edu/abs/2015ApJ...804L..16K},
	doi = {10.1088/2041-8205/804/1/L16},
	urldate = {2026-02-27},
	journal = {The Astrophysical Journal},
	publisher = {IOP},
	author = {Kisaka, Shota and Ioka, Kunihito},
	month = may,
	year = {2015},
	note = {ADS Bibcode: 2015ApJ...804L..16K},
	pages = {L16},
}

@article{gompertz_magnetar_2014,
	title = {Magnetar powered {GRBs}: explaining the extended emission and {X}-ray plateau of short {GRB} light curves},
	volume = {438},
	issn = {0035-8711},
	shorttitle = {Magnetar powered {GRBs}},
	url = {https://ui.adsabs.harvard.edu/abs/2014MNRAS.438..240G},
	doi = {10.1093/mnras/stt2165},
	urldate = {2026-02-27},
	journal = {Monthly Notices of the Royal Astronomical Society},
	publisher = {OUP},
	author = {Gompertz, B. P. and O'Brien, P. T. and Wynn, G. A.},
	month = feb,
	year = {2014},
	note = {ADS Bibcode: 2014MNRAS.438..240G},
	pages = {240--250},
}

@article{nakamura_self-similar_2006,
	title = {Self-similar {Solutions} for the {Interaction} of {Relativistic} {Ejecta} with an {Ambient} {Medium}},
	volume = {645},
	issn = {0004-637X},
	url = {https://iopscience.iop.org/article/10.1086/504025},
	doi = {10.1086/504025},
	language = {en},
	number = {1},
	urldate = {2026-02-19},
	journal = {The Astrophysical Journal},
	publisher = {IOP Publishing},
	author = {Nakamura, Ko and Shigeyama, Toshikazu},
	month = jul,
	year = {2006},
	pages = {431},
}

@misc{fraija_grb250704bep250704a_2026,
	title = {{GRB}{\textasciitilde}{250704B}/{EP250704a} a {Short} {Gamma}-{Ray} {Burst} {Powered} by a {Magnetar}},
	url = {https://arxiv.org/abs/2601.15732v1},
	language = {en},
	urldate = {2026-02-18},
	journal = {arXiv.org},
	author = {Fraija, Nissim and Galvá, Antonio and Kamenetskaia, Boris Betancourt and Dainotti, Maria G.},
	month = jan,
	year = {2026},
}

@article{norris_short_2006,
	title = {Short {Gamma}-{Ray} {Bursts} with {Extended} {Emission}},
	volume = {643},
	issn = {0004-637X},
	url = {https://ui.adsabs.harvard.edu/abs/2006ApJ...643..266N},
	doi = {10.1086/502796},
	urldate = {2026-02-18},
	journal = {The Astrophysical Journal},
	publisher = {IOP},
	author = {Norris, J. P. and Bonnell, J. T.},
	month = may,
	year = {2006},
	note = {ADS Bibcode: 2006ApJ...643..266N},
	pages = {266--275},
}

@article{matsumoto_linking_2020,
	title = {Linking extended and plateau emissions of short gamma-ray bursts},
	volume = {493},
	issn = {0035-8711},
	url = {https://dx.doi.org/10.1093/mnras/staa305},
	doi = {10.1093/mnras/staa305},
	language = {en},
	number = {1},
	urldate = {2026-02-18},
	journal = {Monthly Notices of the Royal Astronomical Society},
	publisher = {Oxford Academic},
	author = {Matsumoto, Tatsuya and Kimura, Shigeo S. and Murase, Kohta and Mészáros, Peter},
	month = mar,
	year = {2020},
	pages = {783--791},
}

@article{kusafuka_ejecta_2025,
	title = {Ejecta width and magnetization reflected in gamma-ray burst early afterglows: implication for reverse shock component and shallow decay phase},
	volume = {536},
	issn = {0035-8711},
	shorttitle = {Ejecta width and magnetization reflected in gamma-ray burst early afterglows},
	url = {https://ui.adsabs.harvard.edu/abs/2025MNRAS.536.1822K},
	doi = {10.1093/mnras/stae2734},
	urldate = {2026-02-06},
	journal = {Monthly Notices of the Royal Astronomical Society},
	publisher = {OUP},
	author = {Kusafuka, Yo and Asano, Katsuaki},
	month = jan,
	year = {2025},
	note = {ADS Bibcode: 2025MNRAS.536.1822K},
	pages = {1822--1837},
}

@article{mroczkowski_conceptual_2025,
	title = {The conceptual design of the 50-meter {Atacama} {Large} {Aperture} {Submillimeter} {Telescope} ({AtLAST})},
	volume = {694},
	issn = {0004-6361},
	url = {https://ui.adsabs.harvard.edu/abs/2025A&A...694A.142M},
	doi = {10.1051/0004-6361/202449786},
	urldate = {2026-02-03},
	journal = {Astronomy and Astrophysics},
	publisher = {EDP},
	author = {Mroczkowski, Tony and Gallardo, Patricio A. and Timpe, Martin and Kiselev, Aleksej and Groh, Manuel and Kaercher, Hans and Reichert, Matthias and Cicone, Claudia and Puddu, Roberto and Dubois-dit-Bonclaude, Pierre and Bok, Daniel and Dahl, Erik and Macintosh, Mike and Dicker, Simon and Viole, Isabelle and Sartori, Sabrina and Valenzuela Venegas, Guillermo Andrés and Zeyringer, Marianne and Niemack, Michael and Poppi, Sergio and Olguin, Rodrigo and Hatziminaoglou, Evanthia and De Breuck, Carlos and Klaassen, Pamela and Montenegro-Montes, Francisco Miguel and Zimmerer, Thomas},
	month = feb,
	year = {2025},
	note = {ADS Bibcode: 2025A\&A...694A.142M},
	pages = {A142},
}

@article{dainotti_discovery_2010,
	title = {Discovery of a {Tight} {Correlation} for {Gamma}-ray {Burst} {Afterglows} with "{Canonical}" {Light} {Curves}},
	volume = {722},
	issn = {0004-637X},
	url = {https://ui.adsabs.harvard.edu/abs/2010ApJ...722L.215D},
	doi = {10.1088/2041-8205/722/2/L215},
	urldate = {2026-02-02},
	journal = {The Astrophysical Journal},
	publisher = {IOP},
	author = {Dainotti, Maria Giovanna and Willingale, Richard and Capozziello, Salvatore and Fabrizio Cardone, Vincenzo and Ostrowski, Michał},
	month = oct,
	year = {2010},
	note = {ADS Bibcode: 2010ApJ...722L.215D},
	pages = {L215--L219},
}

@article{uhm_semi-analytic_2011,
	title = {A {SEMI}-{ANALYTIC} {FORMULATION} {FOR} {RELATIVISTIC} {BLAST} {WAVES} {WITH} {A} {LONG}-{LIVED} {REVERSE} {SHOCK}},
	volume = {733},
	issn = {0004-637X},
	url = {https://doi.org/10.1088/0004-637X/733/2/86},
	doi = {10.1088/0004-637X/733/2/86},
	language = {en},
	number = {2},
	urldate = {2026-01-19},
	journal = {The Astrophysical Journal},
	publisher = {The American Astronomical Society},
	author = {Uhm, Z. Lucas},
	month = may,
	year = {2011},
	pages = {86},
}

@article{kumar_evolution_2003,
	title = {The {Evolution} of a {Structured} {Relativistic} {Jet} and {Gamma}-{Ray} {Burst} {Afterglow} {Light} {Curves}},
	volume = {591},
	issn = {0004-637X},
	url = {https://iopscience.iop.org/article/10.1086/375186},
	doi = {10.1086/375186},
	language = {en},
	number = {2},
	urldate = {2026-01-19},
	journal = {The Astrophysical Journal},
	publisher = {IOP Publishing},
	author = {Kumar, Pawan and Granot, Jonathan},
	month = jul,
	year = {2003},
	pages = {1075},
}

@article{zhang_gamma-ray_2004,
	title = {Gamma-{Ray} {Bursts}: progress, problems \& prospects},
	volume = {19},
	issn = {0217-751X},
	shorttitle = {Gamma-{Ray} {Bursts}},
	url = {https://ui.adsabs.harvard.edu/abs/2004IJMPA..19.2385Z},
	doi = {10.1142/S0217751X0401746X},
	urldate = {2026-01-17},
	journal = {International Journal of Modern Physics A},
	publisher = {WSPC},
	author = {Zhang, Bing and Mészáros, Peter},
	month = jan,
	year = {2004},
	note = {ADS Bibcode: 2004IJMPA..19.2385Z},
	pages = {2385--2472},
}

@article{panaitescu_rings_1998,
	title = {Rings in {Fireball} {Afterglows}},
	volume = {493},
	issn = {0004-637X},
	url = {https://iopscience.iop.org/article/10.1086/311127},
	doi = {10.1086/311127},
	language = {en},
	number = {1},
	urldate = {2026-01-17},
	journal = {The Astrophysical Journal},
	publisher = {IOP Publishing},
	author = {Panaitescu, A. and Mészáros, P.},
	month = jan,
	year = {1998},
	pages = {L31},
}

@article{nakar_early_2004,
	title = {Early afterglow emission from a reverse shock as a diagnostic tool for gamma-ray burst outflows},
	volume = {353},
	issn = {0035-8711},
	url = {https://doi.org/10.1111/j.1365-2966.2004.08099.x},
	doi = {10.1111/j.1365-2966.2004.08099.x},
	number = {2},
	urldate = {2026-01-17},
	journal = {Monthly Notices of the Royal Astronomical Society},
	author = {Nakar, Ehud and Piran, Tsvi},
	month = sep,
	year = {2004},
	pages = {647--653},
}

@book{zhang_physics_2018,
	title = {The {Physics} of {Gamma}-{Ray} {Bursts}},
	url = {https://ui.adsabs.harvard.edu/abs/2018pgrb.book.....Z},
	doi = {10.1017/9781139226530},
	urldate = {2026-01-17},
	author = {Zhang, Bing},
	month = dec,
	year = {2018},
	note = {Publication Title: The Physics of Gamma-Ray Bursts by Bing Zhang. ISBN: 978-1-139-22653-0. Cambridge Univeristy Press
ADS Bibcode: 2018pgrb.book.....Z},
}

@article{sari_impulsive_2000,
	title = {Impulsive and {Varying} {Injection} in {Gamma}-{Ray} {BurstAfterglows}},
	volume = {535},
	issn = {0004-637X},
	url = {https://iopscience.iop.org/article/10.1086/312689},
	doi = {10.1086/312689},
	language = {en},
	number = {1},
	urldate = {2026-01-15},
	journal = {The Astrophysical Journal},
	publisher = {IOP Publishing},
	author = {Sari, Re’em and Mészáros, Peter},
	month = may,
	year = {2000},
	pages = {L33},
}

@article{dai_afterglow_2001,
	title = {Afterglow {Emission} from {Highly} {Collimated} {Jets} with {Flat} {Electron} {Spectra}: {Application} to the {GRB} 010222 {Case}?},
	volume = {558},
	issn = {0004-637X},
	shorttitle = {Afterglow {Emission} from {Highly} {Collimated} {Jets} with {Flat} {Electron} {Spectra}},
	url = {https://iopscience.iop.org/article/10.1086/323566},
	doi = {10.1086/323566},
	language = {en},
	number = {2},
	urldate = {2026-01-15},
	journal = {The Astrophysical Journal},
	publisher = {IOP Publishing},
	author = {Dai, Z. G. and Cheng, K. S.},
	month = aug,
	year = {2001},
	pages = {L109},
}

@article{spitkovsky_particle_2008,
	title = {Particle {Acceleration} in {Relativistic} {Collisionless} {Shocks}: {Fermi} {Process} at {Last}?},
	volume = {682},
	issn = {0004-637X},
	shorttitle = {Particle {Acceleration} in {Relativistic} {Collisionless} {Shocks}},
	url = {https://iopscience.iop.org/article/10.1086/590248},
	doi = {10.1086/590248},
	language = {en},
	number = {1},
	urldate = {2026-01-14},
	journal = {The Astrophysical Journal},
	publisher = {IOP Publishing},
	author = {Spitkovsky, Anatoly},
	month = jul,
	year = {2008},
	pages = {L5},
}

@article{sironi_particle_2009,
	title = {{PARTICLE} {ACCELERATION} {IN} {RELATIVISTIC} {MAGNETIZED} {COLLISIONLESS} {PAIR} {SHOCKS}: {DEPENDENCE} {OF} {SHOCK} {ACCELERATION} {ON} {MAGNETIC} {OBLIQUITY}},
	volume = {698},
	issn = {0004-637X},
	shorttitle = {{PARTICLE} {ACCELERATION} {IN} {RELATIVISTIC} {MAGNETIZED} {COLLISIONLESS} {PAIR} {SHOCKS}},
	url = {https://doi.org/10.1088/0004-637X/698/2/1523},
	doi = {10.1088/0004-637X/698/2/1523},
	language = {en},
	number = {2},
	urldate = {2026-01-14},
	journal = {The Astrophysical Journal},
	publisher = {The American Astronomical Society},
	author = {Sironi, Lorenzo and Spitkovsky, Anatoly},
	month = jun,
	year = {2009},
	pages = {1523},
}

@article{zhang_physical_2006,
	title = {Physical {Processes} {Shaping} {Gamma}-{Ray} {Burst} {X}-{Ray} {Afterglow} {Light} {Curves}: {Theoretical} {Implications} from the {Swift} {X}-{Ray} {Telescope} {Observations}},
	volume = {642},
	issn = {0004-637X},
	shorttitle = {Physical {Processes} {Shaping} {Gamma}-{Ray} {Burst} {X}-{Ray} {Afterglow} {Light} {Curves}},
	url = {https://iopscience.iop.org/article/10.1086/500723},
	doi = {10.1086/500723},
	language = {en},
	number = {1},
	urldate = {2026-01-12},
	journal = {The Astrophysical Journal},
	publisher = {IOP Publishing},
	author = {Zhang, Bing and Fan, Y. Z. and Dyks, Jaroslaw and Kobayashi, Shiho and Mészáros, Peter and Burrows, David N. and Nousek, John A. and Gehrels, Neil},
	month = may,
	year = {2006},
	pages = {354},
}

@article{lei_shallow_2011,
	title = {Shallow {Decay} {Phase} of the {Early} {X}-{Ray} {Afterglow} from {External} {Shock} in a {Wind} {Environment}},
	volume = {28},
	issn = {1741-35400256-307X},
	url = {https://ui.adsabs.harvard.edu/abs/2011ChPhL..28l9801L},
	doi = {10.1088/0256-307X/28/12/129801},
	urldate = {2026-01-12},
	journal = {Chinese Physics Letters},
	publisher = {IOP},
	author = {Lei, Hai-Dong and Wang, Jiu-Zhou and Lü, Jing and Zou, Yuan-Chuan},
	month = dec,
	year = {2011},
	note = {ADS Bibcode: 2011ChPhL..28l9801L},
	pages = {129801},
}

@article{duffell_engine_2015,
	title = {{FROM} {ENGINE} {TO} {AFTERGLOW}: {COLLAPSARS} {NATURALLY} {PRODUCE} {TOP}-{HEAVY} {JETS} {AND} {EARLY}-{TIME} {PLATEAUS} {IN} {GAMMA}-{RAY} {BURST} {AFTERGLOWS}},
	volume = {806},
	issn = {0004-637X},
	shorttitle = {{FROM} {ENGINE} {TO} {AFTERGLOW}},
	url = {https://doi.org/10.1088/0004-637X/806/2/205},
	doi = {10.1088/0004-637X/806/2/205},
	language = {en},
	number = {2},
	urldate = {2026-01-09},
	journal = {The Astrophysical Journal},
	publisher = {The American Astronomical Society},
	author = {Duffell, Paul C. and MacFadyen, Andrew I.},
	month = jun,
	year = {2015},
	pages = {205},
}

@article{kobayashi_onset_2007,
	title = {The {Onset} of {Gamma}-{Ray} {Burst} {Afterglow}},
	volume = {655},
	issn = {0004-637X},
	url = {https://iopscience.iop.org/article/10.1086/510203},
	doi = {10.1086/510203},
	language = {en},
	number = {2},
	urldate = {2026-01-09},
	journal = {The Astrophysical Journal},
	publisher = {IOP Publishing},
	author = {Kobayashi, Shiho and Zhang, Bing},
	month = feb,
	year = {2007},
	pages = {973},
}

@article{leventis_plateau_2014,
	title = {The plateau phase of gamma-ray burst afterglows in the thick-shell scenario},
	volume = {437},
	issn = {0035-8711},
	url = {https://doi.org/10.1093/mnras/stt2055},
	doi = {10.1093/mnras/stt2055},
	number = {3},
	urldate = {2026-01-09},
	journal = {Monthly Notices of the Royal Astronomical Society},
	author = {Leventis, K. and Wijers, R. A. M. J. and van der Horst, A. J.},
	month = jan,
	year = {2014},
	pages = {2448--2460},
}

@article{eichler_case_2006,
	title = {The {Case} for {Anisotropic} {Afterglow} {Efficiency} within {Gamma}-{Ray} {Burst} {Jets}},
	volume = {641},
	issn = {0004-637X},
	url = {https://iopscience.iop.org/article/10.1086/503667},
	doi = {10.1086/503667},
	language = {en},
	number = {1},
	urldate = {2026-01-09},
	journal = {The Astrophysical Journal},
	publisher = {IOP Publishing},
	author = {Eichler, David and Granot, Jonathan},
	month = mar,
	year = {2006},
	pages = {L5},
}

@article{beniamini_observational_2019,
	title = {Observational constraints on the structure of gamma-ray burst jets},
	volume = {482},
	issn = {0035-8711},
	url = {https://doi.org/10.1093/mnras/sty3110},
	doi = {10.1093/mnras/sty3110},
	number = {4},
	urldate = {2026-01-09},
	journal = {Monthly Notices of the Royal Astronomical Society},
	author = {Beniamini, Paz and Nakar, Ehud},
	month = feb,
	year = {2019},
	pages = {5430--5440},
}

@article{kobayashi_ultraefficient_2001,
	title = {Ultraefficient {Internal} {Shocks}},
	volume = {551},
	issn = {0004-637X},
	url = {https://iopscience.iop.org/article/10.1086/320249},
	doi = {10.1086/320249},
	language = {en},
	number = {2},
	urldate = {2026-01-09},
	journal = {The Astrophysical Journal},
	publisher = {IOP Publishing},
	author = {Kobayashi, Shiho and Sari, Re’em},
	month = apr,
	year = {2001},
	pages = {934},
}

@article{yu_shallow_2007,
	title = {Shallow decay phase of {GRB} {X}-ray afterglows from relativistic wind bubbles},
	volume = {470},
	copyright = {© ESO, 2007},
	issn = {0004-6361, 1432-0746},
	url = {https://www.aanda.org/articles/aa/abs/2007/28/aa7053-07/aa7053-07.html},
	doi = {10.1051/0004-6361:20077053},
	language = {en},
	number = {1},
	urldate = {2026-01-09},
	journal = {Astronomy \& Astrophysics},
	publisher = {EDP Sciences},
	author = {Yu, Y. W. and Dai, Z. G.},
	month = jul,
	year = {2007},
	pages = {119--122},
}

@article{beniamini_x-ray_2020,
	title = {X-ray plateaus in gamma-ray bursts' light curves from jets viewed slightly off-axis},
	volume = {492},
	issn = {0035-8711},
	url = {https://ui.adsabs.harvard.edu/abs/2020MNRAS.492.2847B},
	doi = {10.1093/mnras/staa070},
	urldate = {2026-01-08},
	journal = {Monthly Notices of the Royal Astronomical Society},
	publisher = {OUP},
	author = {Beniamini, Paz and Duque, Raphaël and Daigne, Frédéric and Mochkovitch, Robert},
	month = feb,
	year = {2020},
	note = {ADS Bibcode: 2020MNRAS.492.2847B},
	pages = {2847--2857},
}

@article{zhang_gamma-ray_2002,
	title = {Gamma-{Ray} {Bursts} with {Continuous} {Energy} {Injection} and {Their} {Afterglow} {Signature}},
	volume = {566},
	issn = {0004-637X},
	url = {https://iopscience.iop.org/article/10.1086/338247},
	doi = {10.1086/338247},
	language = {en},
	number = {2},
	urldate = {2026-01-08},
	journal = {The Astrophysical Journal},
	publisher = {IOP Publishing},
	author = {Zhang, Bing and Mészáros, Peter},
	month = feb,
	year = {2002},
	pages = {712},
}

@article{dallosso_gamma-ray_2011,
	title = {Gamma-ray bursts afterglows with energy injection from a spinning down neutron star},
	volume = {526},
	issn = {0004-6361},
	url = {https://ui.adsabs.harvard.edu/abs/2011A&A...526A.121D},
	doi = {10.1051/0004-6361/201014168},
	urldate = {2026-01-08},
	journal = {Astronomy and Astrophysics},
	publisher = {EDP},
	author = {Dall'Osso, S. and Stratta, G. and Guetta, D. and Covino, S. and De Cesare, G. and Stella, L.},
	month = feb,
	year = {2011},
	note = {ADS Bibcode: 2011A\&A...526A.121D},
	pages = {A121},
}

@article{panaitescu_evidence_2006,
	title = {Evidence for chromatic {X}-ray light-curve breaks in {Swift} gamma-ray burst afterglows and their theoretical implications},
	volume = {369},
	issn = {0035-8711},
	url = {https://ui.adsabs.harvard.edu/abs/2006MNRAS.369.2059P},
	doi = {10.1111/j.1365-2966.2006.10453.x},
	urldate = {2026-01-08},
	journal = {Monthly Notices of the Royal Astronomical Society},
	publisher = {OUP},
	author = {Panaitescu, A. and Mészáros, P. and Burrows, D. and Nousek, J. and Gehrels, N. and O'Brien, P. and Willingale, R.},
	month = jul,
	year = {2006},
	note = {ADS Bibcode: 2006MNRAS.369.2059P},
	pages = {2059--2064},
}

@article{ioka_efficiency_2006,
	title = {Efficiency crisis of swift gamma-ray bursts with shallow {X}-ray afterglows: prior activity or time-dependent microphysics?},
	volume = {458},
	issn = {0004-6361},
	shorttitle = {Efficiency crisis of swift gamma-ray bursts with shallow {X}-ray afterglows},
	url = {https://ui.adsabs.harvard.edu/abs/2006A&A...458....7I},
	doi = {10.1051/0004-6361:20064939},
	urldate = {2026-01-08},
	journal = {Astronomy and Astrophysics},
	publisher = {EDP},
	author = {Ioka, K. and Toma, K. and Yamazaki, R. and Nakamura, T.},
	month = oct,
	year = {2006},
	note = {ADS Bibcode: 2006A\&A...458....7I},
	pages = {7--12},
}

@article{granot_distribution_2006,
	title = {Distribution of gamma-ray burst ejecta energy with {Lorentz} factor},
	volume = {366},
	issn = {1745-3925},
	url = {https://doi.org/10.1111/j.1745-3933.2005.00121.x},
	doi = {10.1111/j.1745-3933.2005.00121.x},
	number = {1},
	urldate = {2026-01-08},
	journal = {Monthly Notices of the Royal Astronomical Society: Letters},
	author = {Granot, Jonathan and Kumar, Pawan},
	month = feb,
	year = {2006},
	pages = {L13--L16},
}

@misc{li_statistical_2026,
	title = {Statistical analysis of multi-band plateaus in gamma-ray burst afterglows},
	url = {https://ui.adsabs.harvard.edu/abs/2026arXiv260101586L},
	urldate = {2026-01-06},
	publisher = {arXiv},
	author = {Li, Xiao-Yan and Liu, Tong and Huang, Bao-Quan and Deng, Chen},
	month = jan,
	year = {2026},
	note = {ADS Bibcode: 2026arXiv260101586L},
}

@article{waxman_neutrino_2000,
	title = {Neutrino {Afterglow} from {Gamma}-{Ray} {Bursts}: {\textasciitilde}1018 {EV}},
	volume = {541},
	issn = {0004-637X},
	shorttitle = {Neutrino {Afterglow} from {Gamma}-{Ray} {Bursts}},
	url = {https://ui.adsabs.harvard.edu/abs/2000ApJ...541..707W},
	doi = {10.1086/309462},
	urldate = {2025-12-29},
	journal = {The Astrophysical Journal},
	publisher = {IOP},
	author = {Waxman, Eli and Bahcall, John N.},
	month = oct,
	year = {2000},
	note = {ADS Bibcode: 2000ApJ...541..707W},
	pages = {707--711},
}

@article{waxman_gamma-ray--burst_1997,
	title = {Gamma-{Ray}--{Burst} {Afterglow}: {Supporting} the {Cosmological} {Fireball} {Model}, {Constraining} {Parameters}, and {Making} {Predictions}},
	volume = {485},
	issn = {0004-637X},
	shorttitle = {Gamma-{Ray}--{Burst} {Afterglow}},
	url = {https://ui.adsabs.harvard.edu/abs/1997ApJ...485L...5W},
	doi = {10.1086/310809},
	urldate = {2025-12-29},
	journal = {The Astrophysical Journal},
	publisher = {IOP},
	author = {Waxman, Eli},
	month = aug,
	year = {1997},
	note = {ADS Bibcode: 1997ApJ...485L...5W},
	pages = {L5--L8},
}

@article{wygoda_energy_2016,
	title = {The {Energy} {Budget} of {GRBs} {Based} on a {Large} {Sample} of {Prompt} and {Afterglow} {Observations}},
	volume = {824},
	issn = {0004-637X},
	url = {https://ui.adsabs.harvard.edu/abs/2016ApJ...824..127W},
	doi = {10.3847/0004-637X/824/2/127},
	urldate = {2025-12-29},
	journal = {The Astrophysical Journal},
	publisher = {IOP},
	author = {Wygoda, N. and Guetta, D. and Mandich, M. A. and Waxman, E.},
	month = jun,
	year = {2016},
	note = {ADS Bibcode: 2016ApJ...824..127W},
	pages = {127},
}

@article{zhang_semi-analytical_2022,
	title = {A semi-analytical solution to the forward-reverse shock hydrodynamics of the gamma-ray burst afterglow},
	volume = {513},
	issn = {0035-8711},
	url = {https://ui.adsabs.harvard.edu/abs/2022MNRAS.513.4887Z},
	doi = {10.1093/mnras/stac1198},
	urldate = {2025-12-28},
	journal = {Monthly Notices of the Royal Astronomical Society},
	publisher = {OUP},
	author = {Zhang, Ze-Lin and Liu, Ruo-Yu and Geng, Jin-Jun and Wu, Xue-Feng and Wang, Xiang-Yu},
	month = jul,
	year = {2022},
	note = {ADS Bibcode: 2022MNRAS.513.4887Z},
	pages = {4887--4898},
}

@article{oganesyan_structured_2020,
	title = {Structured {Jets} and {X}-{Ray} {Plateaus} in {Gamma}-{Ray} {Burst} {Phenomena}},
	volume = {893},
	issn = {0004-637X},
	url = {https://doi.org/10.3847/1538-4357/ab8221},
	doi = {10.3847/1538-4357/ab8221},
	language = {en},
	number = {2},
	urldate = {2025-12-12},
	journal = {The Astrophysical Journal},
	publisher = {The American Astronomical Society},
	author = {Oganesyan, Gor and Ascenzi, Stefano and Branchesi, Marica and Salafia, Om Sharan and Dall’Osso, Simone and Ghirlanda, Giancarlo},
	month = apr,
	year = {2020},
	pages = {88},
}

@article{metzger_protomagnetar_2011,
	title = {The protomagnetar model for gamma-ray bursts},
	volume = {413},
	issn = {0035-8711},
	url = {https://doi.org/10.1111/j.1365-2966.2011.18280.x},
	doi = {10.1111/j.1365-2966.2011.18280.x},
	number = {3},
	urldate = {2025-12-12},
	journal = {Monthly Notices of the Royal Astronomical Society},
	author = {Metzger, B. D. and Giannios, D. and Thompson, T. A. and Bucciantini, N. and Quataert, E.},
	month = may,
	year = {2011},
	pages = {2031--2056},
}

@article{shen_coasting_2011,
	title = {{COASTING} {EXTERNAL} {SHOCK} {IN} {WIND} {MEDIUM}: {AN} {ORIGIN} {FOR} {THE} {X}-{RAY} {PLATEAU} {DECAY} {COMPONENT} {IN} {SWIFT} {GAMMA}-{RAY} {BURST} {AFTERGLOWS}},
	volume = {744},
	issn = {0004-637X},
	shorttitle = {{COASTING} {EXTERNAL} {SHOCK} {IN} {WIND} {MEDIUM}},
	url = {https://doi.org/10.1088/0004-637X/744/1/36},
	doi = {10.1088/0004-637X/744/1/36},
	language = {en},
	number = {1},
	urldate = {2025-12-12},
	journal = {The Astrophysical Journal},
	publisher = {The American Astronomical Society},
	author = {Shen, Rongfeng and Matzner, Christopher D.},
	month = dec,
	year = {2011},
	pages = {36},
}

@article{rees_relativistic_1992,
	title = {Relativistic fireballs - {Energy} conversion and time-scales.},
	volume = {258},
	issn = {0035-8711},
	url = {https://ui.adsabs.harvard.edu/abs/1992MNRAS.258P..41R},
	doi = {10.1093/mnras/258.1.41P},
	urldate = {2025-12-09},
	journal = {Monthly Notices of the Royal Astronomical Society},
	publisher = {OUP},
	author = {Rees, M. J. and Meszaros, P.},
	month = sep,
	year = {1992},
	note = {ADS Bibcode: 1992MNRAS.258P..41R},
	pages = {41},
}

@article{sari_hydrodynamics_1997,
	title = {Hydrodynamics of {Gamma}-{Ray} {Burst} {Afterglow}},
	volume = {489},
	issn = {0004-637X},
	url = {https://ui.adsabs.harvard.edu/abs/1997ApJ...489L..37S},
	doi = {10.1086/310957},
	urldate = {2025-12-04},
	journal = {The Astrophysical Journal},
	publisher = {IOP},
	author = {Sari, Re'em},
	month = nov,
	year = {1997},
	note = {ADS Bibcode: 1997ApJ...489L..37S},
	pages = {L37--L40},
}

@article{kobayashi_hydrodynamics_1999,
	title = {Hydrodynamics of a {Relativistic} {Fireball}: {The} {Complete} {Evolution}},
	volume = {513},
	issn = {0004-637X},
	shorttitle = {Hydrodynamics of a {Relativistic} {Fireball}},
	url = {https://iopscience.iop.org/article/10.1086/306868/meta},
	doi = {10.1086/306868},
	language = {en},
	number = {2},
	urldate = {2025-12-03},
	journal = {The Astrophysical Journal},
	publisher = {IOP Publishing},
	author = {Kobayashi, Shiho and Piran, Tsvi and Sari, Re'em},
	month = mar,
	year = {1999},
	pages = {669},
}

@book{synge_relativistic_1957,
	address = {Amsterdam, New York},
	title = {The relativistic gas.},
	language = {eng},
	publisher = {North-Holland Pub. Co.},
	author = {Synge, J. L.},
	year = {1957},
	note = {Open Library ID: OL6219379M},
}

@article{sari_hydrodynamic_1995,
	title = {Hydrodynamic {Timescales} and {Temporal} {Structure} of {Gamma}-{Ray} {Bursts}},
	volume = {455},
	issn = {0004-637X},
	url = {https://ui.adsabs.harvard.edu/abs/1995ApJ...455L.143S},
	doi = {10.1086/309835},
	urldate = {2025-11-26},
	journal = {The Astrophysical Journal},
	publisher = {IOP},
	author = {Sari, Re'em and Piran, Tsvi},
	month = dec,
	year = {1995},
	note = {ADS Bibcode: 1995ApJ...455L.143S},
	pages = {L143},
}

@article{genet_can_2007,
	title = {Can the early {X}-ray afterglow of gamma-ray bursts be explained by a contribution from the reverse shock?},
	volume = {381},
	issn = {0035-8711},
	url = {https://ui.adsabs.harvard.edu/abs/2007MNRAS.381..732G},
	doi = {10.1111/j.1365-2966.2007.12243.x},
	urldate = {2025-11-24},
	journal = {Monthly Notices of the Royal Astronomical Society},
	publisher = {OUP},
	author = {Genet, F. and Daigne, F. and Mochkovitch, R.},
	month = oct,
	year = {2007},
	note = {ADS Bibcode: 2007MNRAS.381..732G},
	pages = {732--740},
}

@article{nousek_evidence_2006,
	title = {Evidence for a {Canonical} {Gamma}-{Ray} {Burst} {Afterglow} {Light} {Curve} in the {Swift} {XRT} {Data}},
	volume = {642},
	issn = {0004-637X},
	url = {https://ui.adsabs.harvard.edu/abs/2006ApJ...642..389N},
	doi = {10.1086/500724},
	urldate = {2025-11-24},
	journal = {The Astrophysical Journal},
	publisher = {IOP},
	author = {Nousek, J. A. and Kouveliotou, C. and Grupe, D. and Page, K. L. and Granot, J. and Ramirez-Ruiz, E. and Patel, S. K. and Burrows, D. N. and Mangano, V. and Barthelmy, S. and Beardmore, A. P. and Campana, S. and Capalbi, M. and Chincarini, G. and Cusumano, G. and Falcone, A. D. and Gehrels, N. and Giommi, P. and Goad, M. R. and Godet, O. and Hurkett, C. P. and Kennea, J. A. and Moretti, A. and O'Brien, P. T. and Osborne, J. P. and Romano, P. and Tagliaferri, G. and Wells, A. A.},
	month = may,
	year = {2006},
	note = {ADS Bibcode: 2006ApJ...642..389N},
	pages = {389--400},
}

@article{meszaros_optical_1997,
	title = {Optical and {Long}-{Wavelength} {Afterglow} from {Gamma}-{Ray} {Bursts}},
	volume = {476},
	issn = {0004-637X},
	url = {https://ui.adsabs.harvard.edu/abs/1997ApJ...476..232M},
	doi = {10.1086/303625},
	urldate = {2025-11-24},
	journal = {The Astrophysical Journal},
	publisher = {IOP},
	author = {Mészáros, P. and Rees, M. J.},
	month = feb,
	year = {1997},
	note = {ADS Bibcode: 1997ApJ...476..232M},
	pages = {232--237},
}

@article{rees_refreshed_1998,
	title = {Refreshed {Shocks} and {Afterglow} {Longevity} in {Gamma}-{Ray} {Bursts}},
	volume = {496},
	issn = {0004-637X},
	url = {https://ui.adsabs.harvard.edu/abs/1998ApJ...496L...1R},
	doi = {10.1086/311244},
	urldate = {2025-11-24},
	journal = {The Astrophysical Journal},
	publisher = {IOP},
	author = {Rees, M. J. and Mészáros, P.},
	month = mar,
	year = {1998},
	note = {ADS Bibcode: 1998ApJ...496L...1R},
	pages = {L1--L4},
}

@article{uhm_mechanism_2007,
	title = {On the {Mechanism} of {Gamma}-{Ray} {Burst} {Afterglows}},
	volume = {665},
	issn = {0004-637X},
	url = {https://ui.adsabs.harvard.edu/abs/2007ApJ...665L..93U},
	doi = {10.1086/519837},
	urldate = {2025-11-24},
	journal = {The Astrophysical Journal},
	publisher = {IOP},
	author = {Uhm, Z. Lucas and Beloborodov, Andrei M.},
	month = aug,
	year = {2007},
	note = {ADS Bibcode: 2007ApJ...665L..93U},
	pages = {L93--L96},
}

@article{troja_swift_2007,
	title = {Swift {Observations} of {GRB} 070110: {An} {Extraordinary} {X}-{Ray} {Afterglow} {Powered} by the {Central} {Engine}},
	volume = {665},
	issn = {0004-637X},
	shorttitle = {Swift {Observations} of {GRB} 070110},
	url = {https://ui.adsabs.harvard.edu/abs/2007ApJ...665..599T},
	doi = {10.1086/519450},
	urldate = {2025-11-24},
	journal = {The Astrophysical Journal},
	publisher = {IOP},
	author = {Troja, E. and Cusumano, G. and O'Brien, P. T. and Zhang, B. and Sbarufatti, B. and Mangano, V. and Willingale, R. and Chincarini, G. and Osborne, J. P. and Marshall, F. E. and Burrows, D. N. and Campana, S. and Gehrels, N. and Guidorzi, C. and Krimm, H. A. and La Parola, V. and Liang, E. W. and Mineo, T. and Moretti, A. and Page, K. L. and Romano, P. and Tagliaferri, G. and Zhang, B. B. and Page, M. J. and Schady, P.},
	month = aug,
	year = {2007},
	note = {ADS Bibcode: 2007ApJ...665..599T},
	pages = {599--607},
}

@article{liang_comprehensive_2007,
	title = {A {Comprehensive} {Analysis} of {Swift} {XRT} {Data}. {II}. {Diverse} {Physical} {Origins} of the {Shallow} {Decay} {Segment}},
	volume = {670},
	issn = {0004-637X},
	url = {https://ui.adsabs.harvard.edu/abs/2007ApJ...670..565L},
	doi = {10.1086/521870},
	urldate = {2025-11-24},
	journal = {The Astrophysical Journal},
	publisher = {IOP},
	author = {Liang, En-Wei and Zhang, Bin-Bin and Zhang, Bing},
	month = nov,
	year = {2007},
	note = {ADS Bibcode: 2007ApJ...670..565L},
	pages = {565--583},
}

@article{ror_investigating_2025,
	title = {Investigating temporal features in {Swift} {GRB} afterglows: a comparative study of {UVOT} and {XRT} data},
	volume = {543},
	issn = {0035-8711},
	shorttitle = {Investigating temporal features in {Swift} {GRB} afterglows},
	url = {https://ui.adsabs.harvard.edu/abs/2025MNRAS.543.2404R},
	doi = {10.1093/mnras/staf1514},
	urldate = {2025-11-24},
	journal = {Monthly Notices of the Royal Astronomical Society},
	publisher = {OUP},
	author = {Ror, Amit K. and Pandey, S. B. and Oates, S. R. and Gupta, R. and Aryan, A. and Castro-Tirado, A. J. and Kumar, Sudhir},
	month = nov,
	year = {2025},
	note = {ADS Bibcode: 2025MNRAS.543.2404R},
	pages = {2404--2441},
}

@article{burrows_swift_2005,
	title = {The {Swift} {X}-{Ray} {Telescope}},
	volume = {120},
	issn = {0038-6308},
	url = {https://ui.adsabs.harvard.edu/abs/2005SSRv..120..165B},
	doi = {10.1007/s11214-005-5097-2},
	urldate = {2025-11-24},
	journal = {Space Science Reviews},
	author = {Burrows, David N. and Hill, J. E. and Nousek, J. A. and Kennea, J. A. and Wells, A. and Osborne, J. P. and Abbey, A. F. and Beardmore, A. and Mukerjee, K. and Short, A. D. T. and Chincarini, G. and Campana, S. and Citterio, O. and Moretti, A. and Pagani, C. and Tagliaferri, G. and Giommi, P. and Capalbi, M. and Tamburelli, F. and Angelini, L. and Cusumano, G. and Bräuninger, H. W. and Burkert, W. and Hartner, G. D.},
	month = oct,
	year = {2005},
	note = {ADS Bibcode: 2005SSRv..120..165B},
	pages = {165--195},
}

@article{gehrels_swift_2004,
	title = {The {Swift} {Gamma}-{Ray} {Burst} {Mission}},
	volume = {611},
	issn = {0004-637X},
	url = {https://ui.adsabs.harvard.edu/abs/2004ApJ...611.1005G},
	doi = {10.1086/422091},
	urldate = {2025-11-24},
	journal = {The Astrophysical Journal},
	publisher = {IOP},
	author = {Gehrels, N. and Chincarini, G. and Giommi, P. and Mason, K. O. and Nousek, J. A. and Wells, A. A. and White, N. E. and Barthelmy, S. D. and Burrows, D. N. and Cominsky, L. R. and Hurley, K. C. and Marshall, F. E. and Mészáros, P. and Roming, P. W. A. and Angelini, L. and Barbier, L. M. and Belloni, T. and Campana, S. and Caraveo, P. A. and Chester, M. M. and Citterio, O. and Cline, T. L. and Cropper, M. S. and Cummings, J. R. and Dean, A. J. and Feigelson, E. D. and Fenimore, E. E. and Frail, D. A. and Fruchter, A. S. and Garmire, G. P. and Gendreau, K. and Ghisellini, G. and Greiner, J. and Hill, J. E. and Hunsberger, S. D. and Krimm, H. A. and Kulkarni, S. R. and Kumar, P. and Lebrun, F. and Lloyd-Ronning, N. M. and Markwardt, C. B. and Mattson, B. J. and Mushotzky, R. F. and Norris, J. P. and Osborne, J. and Paczynski, B. and Palmer, D. M. and Park, H.-S. and Parsons, A. M. and Paul, J. and Rees, M. J. and Reynolds, C. S. and Rhoads, J. E. and Sasseen, T. P. and Schaefer, B. E. and Short, A. T. and Smale, A. P. and Smith, I. A. and Stella, L. and Tagliaferri, G. and Takahashi, T. and Tashiro, M. and Townsley, L. K. and Tueller, J. and Turner, M. J. L. and Vietri, M. and Voges, W. and Ward, M. J. and Willingale, R. and Zerbi, F. M. and Zhang, W. W.},
	month = aug,
	year = {2004},
	note = {ADS Bibcode: 2004ApJ...611.1005G},
	pages = {1005--1020},
}

@article{granot_shape_2002,
	title = {The {Shape} of {Spectral} {Breaks} in {Gamma}-{Ray} {Burst} {Afterglows}},
	volume = {568},
	issn = {0004-637X},
	url = {https://ui.adsabs.harvard.edu/abs/2002ApJ...568..820G},
	doi = {10.1086/338966},
	urldate = {2025-11-24},
	journal = {The Astrophysical Journal},
	publisher = {IOP},
	author = {Granot, Jonathan and Sari, Re'em},
	month = apr,
	year = {2002},
	note = {ADS Bibcode: 2002ApJ...568..820G},
	pages = {820--829},
}

@article{berger_short-duration_2014,
	title = {Short-{Duration} {Gamma}-{Ray} {Bursts}},
	volume = {52},
	issn = {0066-4146},
	url = {https://ui.adsabs.harvard.edu/abs/2014ARA&A..52...43B},
	doi = {10.1146/annurev-astro-081913-035926},
	urldate = {2025-11-24},
	journal = {Annual Review of Astronomy and Astrophysics},
	author = {Berger, Edo},
	month = aug,
	year = {2014},
	note = {ADS Bibcode: 2014ARA\&A..52...43B},
	pages = {43--105},
}

@article{kumar_physics_2015,
	title = {The physics of gamma-ray bursts \& relativistic jets},
	volume = {561},
	issn = {0370-1573},
	url = {https://ui.adsabs.harvard.edu/abs/2015PhR...561....1K},
	doi = {10.1016/j.physrep.2014.09.008},
	urldate = {2025-11-24},
	journal = {Physics Reports},
	publisher = {Elsevier},
	author = {Kumar, Pawan and Zhang, Bing},
	month = feb,
	year = {2015},
	note = {ADS Bibcode: 2015PhR...561....1K},
	pages = {1--109},
}

@article{meszaros_gamma-ray_2006,
	title = {Gamma-ray bursts},
	volume = {69},
	issn = {0034-4885},
	url = {https://ui.adsabs.harvard.edu/abs/2006RPPh...69.2259M},
	doi = {10.1088/0034-4885/69/8/R01},
	urldate = {2025-11-24},
	journal = {Reports on Progress in Physics},
	publisher = {IOP},
	author = {Mészáros, P.},
	month = aug,
	year = {2006},
	note = {ADS Bibcode: 2006RPPh...69.2259M},
	pages = {2259--2321},
}

@article{ronchini_combined_2023,
	title = {Combined {X}-ray and optical analysis to probe the origin of the plateau emission in γ-ray burst afterglows},
	volume = {675},
	copyright = {© The Authors 2023},
	issn = {0004-6361, 1432-0746},
	url = {https://www.aanda.org/articles/aa/abs/2023/07/aa45348-22/aa45348-22.html},
	doi = {10.1051/0004-6361/202245348},
	language = {en},
	urldate = {2025-11-20},
	journal = {Astronomy \& Astrophysics},
	publisher = {EDP Sciences},
	author = {Ronchini, S. and Stratta, G. and Rossi, A. and Kann, D. A. and Oganeysan, G. and Dall’Osso, S. and Branchesi, M. and Cesare, G. De},
	month = jul,
	year = {2023},
	pages = {A117},
}

@misc{swain_grb_2025,
	title = {{GRB} {250704B}: {An} {Off}-axis {Short} {GRB} with a {Long}-{Lived} {Afterglow} {Plateau}},
	shorttitle = {{GRB} {250704B}},
	url = {http://arxiv.org/abs/2509.02769},
	doi = {10.48550/arXiv.2509.02769},
	urldate = {2025-11-17},
	publisher = {arXiv},
	author = {Swain, Vishwajeet and Ahumada, Tomás and Patil, Sameer K. and Wagh, Yogesh and Bhalerao, Varun and Nakar, Ehud and Kasliwal, Mansi and Hall, Xander J. and Busmann, Malte and Anand, Shreya and Karambelkar, Viraj and Andreoni, Igor and Anupama, G. C. and Arya, Anuraag and Balasubramanian, Arvind and Barway, Sudhanshu and Carney, Jonathan and Coughlin, Michael and Eappachen, Deepak and Freeburn, James and Gruen, Daniel and Mohan, Tanishk and O'Connor, Brendan and Palmese, Antonella and Pathak, Utkarsh and Sahu, D. K. and Saikia, Aditya Pawan and Sarin, Nikhil and Srinivasaragavan, Gokul and Tanenia, Hitesh},
	month = sep,
	year = {2025},
	note = {arXiv:2509.02769 [astro-ph]},
}

@article{dainotti_timeluminosity_2008,
	title = {A time–luminosity correlation for γ-ray bursts in the {X}-rays},
	volume = {391},
	issn = {1745-3925},
	url = {https://doi.org/10.1111/j.1745-3933.2008.00560.x},
	doi = {10.1111/j.1745-3933.2008.00560.x},
	number = {1},
	urldate = {2025-11-11},
	journal = {Monthly Notices of the Royal Astronomical Society: Letters},
	author = {Dainotti, M. G. and Cardone, V. F. and Capozziello, S.},
	month = nov,
	year = {2008},
	pages = {L79--L83},
}

@article{granot_diagnosing_2006,
	title = {Diagnosing the {Outflow} from the {SGR} 1806–20 {Giant} {Flare} with {Radio} {Observations}},
	volume = {638},
	issn = {0004-637X},
	url = {https://iopscience.iop.org/article/10.1086/497680/meta},
	doi = {10.1086/497680},
	language = {en},
	number = {1},
	urldate = {2025-08-06},
	journal = {The Astrophysical Journal},
	publisher = {IOP Publishing},
	author = {Granot, J. and Ramirez-Ruiz, E. and Taylor, G. B. and Eichler, D. and Lyubarsky, Y. E. and Wijers, R. a. M. J. and Gaensler, B. M. and Gelfand, J. D. and Kouveliotou, C.},
	month = feb,
	year = {2006},
	pages = {391},
}

@article{sadeh_nonthermal_2025,
	title = {The {Nonthermal} {Emission} {Following} {GW170817} is {Consistent} with a {Conical} {Radially} {Stratified} {Outflow} with {Initial} {Lorentz} {Factor} ≲10},
	volume = {987},
	issn = {0004-637X},
	url = {https://dx.doi.org/10.3847/1538-4357/ade150},
	doi = {10.3847/1538-4357/ade150},
	language = {en},
	number = {2},
	urldate = {2025-07-09},
	journal = {The Astrophysical Journal},
	publisher = {The American Astronomical Society},
	author = {Sadeh, Gilad and Waxman, Eli},
	month = jul,
	year = {2025},
	pages = {178},
}

@article{beniamini_afterglow_2020,
	title = {Afterglow light curves from misaligned structured jets},
	volume = {493},
	issn = {0035-8711},
	url = {https://ui.adsabs.harvard.edu/abs/2020MNRAS.493.3521B},
	doi = {10.1093/mnras/staa538},
	urldate = {2025-05-04},
	journal = {Monthly Notices of the Royal Astronomical Society},
	publisher = {OUP},
	author = {Beniamini, Paz and Granot, Jonathan and Gill, Ramandeep},
	month = apr,
	year = {2020},
	note = {ADS Bibcode: 2020MNRAS.493.3521B},
	pages = {3521--3534},
}

@article{sadeh_late-time_2024,
	title = {Late-time non-thermal emission from mildly relativistic tidal ejecta of compact objects merger},
	volume = {535},
	issn = {0035-8711},
	url = {https://doi.org/10.1093/mnras/stae2561},
	doi = {10.1093/mnras/stae2561},
	number = {4},
	urldate = {2024-12-08},
	journal = {Monthly Notices of the Royal Astronomical Society},
	author = {Sadeh, Gilad},
	month = dec,
	year = {2024},
	pages = {3252--3261},
}

@article{dereli-begue_wind_2022,
	title = {A wind environment and {Lorentz} factors of tens explain gamma-ray bursts {X}-ray plateau},
	volume = {13},
	copyright = {2022 The Author(s)},
	issn = {2041-1723},
	url = {https://www.nature.com/articles/s41467-022-32881-1},
	doi = {10.1038/s41467-022-32881-1},
	language = {en},
	number = {1},
	urldate = {2024-10-10},
	journal = {Nature Communications},
	publisher = {Nature Publishing Group},
	author = {Dereli-Bégué, Hüsne and Pe’er, Asaf and Ryde, Felix and Oates, Samantha R. and Zhang, Bing and Dainotti, Maria G.},
	month = sep,
	year = {2022},
	pages = {5611},
}

@article{lithwick_lower_2001,
	title = {Lower {Limits} on {Lorentz} {Factors} in {Gamma}-{Ray} {Bursts}},
	volume = {555},
	issn = {0004-637X},
	url = {https://ui.adsabs.harvard.edu/abs/2001ApJ...555..540L},
	doi = {10.1086/321455},
	urldate = {2024-08-26},
	journal = {The Astrophysical Journal},
	publisher = {IOP},
	author = {Lithwick, Yoram and Sari, Re'em},
	month = jul,
	year = {2001},
	note = {ADS Bibcode: 2001ApJ...555..540L},
	pages = {540--545},
}

@article{sari_observed_1998,
	title = {The {Observed} {Size} and {Shape} of {Gamma}-{Ray} {Burst} {Afterglow}},
	volume = {494},
	issn = {0004-637X},
	url = {https://ui.adsabs.harvard.edu/abs/1998ApJ...494L..49S},
	doi = {10.1086/311160},
	urldate = {2024-05-22},
	journal = {The Astrophysical Journal},
	publisher = {IOP},
	author = {Sari, Re'em},
	month = feb,
	year = {1998},
	note = {ADS Bibcode: 1998ApJ...494L..49S},
	pages = {L49--L52},
}

@article{keshet_analytical_2006,
	title = {Analytical {Study} of {Diffusive} {Relativistic} {Shock} {Acceleration}},
	volume = {97},
	issn = {0031-9007},
	url = {https://ui.adsabs.harvard.edu/abs/2006PhRvL..97v1104K},
	doi = {10.1103/PhysRevLett.97.221104},
	urldate = {2024-05-20},
	journal = {Physical Review Letters},
	publisher = {APS},
	author = {Keshet, Uri},
	month = dec,
	year = {2006},
	note = {ADS Bibcode: 2006PhRvL..97v1104K},
	pages = {221104},
}

@article{mignone_equation_2007,
	title = {Equation of state in relativistic magnetohydrodynamics: variable versus constant adiabatic index},
	volume = {378},
	issn = {0035-8711},
	shorttitle = {Equation of state in relativistic magnetohydrodynamics},
	url = {https://ui.adsabs.harvard.edu/abs/2007MNRAS.378.1118M},
	doi = {10.1111/j.1365-2966.2007.11849.x},
	urldate = {2024-04-16},
	journal = {Monthly Notices of the Royal Astronomical Society},
	publisher = {OUP},
	author = {Mignone, A. and McKinney, Jonathan C.},
	month = jul,
	year = {2007},
	note = {ADS Bibcode: 2007MNRAS.378.1118M},
	pages = {1118--1130},
}

@article{pohl_pic_2020,
	title = {{PIC} simulation methods for cosmic radiation and plasma instabilities},
	volume = {111},
	issn = {0146-6410},
	url = {https://www.sciencedirect.com/science/article/pii/S0146641019300869},
	doi = {10.1016/j.ppnp.2019.103751},
	urldate = {2023-11-06},
	journal = {Progress in Particle and Nuclear Physics},
	author = {Pohl, M. and Hoshino, M. and Niemiec, J.},
	month = mar,
	year = {2020},
	pages = {103751},
}

@article{sadeh_non-thermal_2023,
	title = {Non-thermal emission from mildly relativistic dynamical ejecta of neutron star mergers},
	volume = {518},
	issn = {0035-8711},
	url = {https://doi.org/10.1093/mnras/stac3260},
	doi = {10.1093/mnras/stac3260},
	number = {2},
	urldate = {2023-09-11},
	journal = {Monthly Notices of the Royal Astronomical Society},
	author = {Sadeh, Gilad and Guttman, Or and Waxman, Eli},
	month = jan,
	year = {2023},
	pages = {2102--2112},
}

@book{rybicki_radiative_1979,
	title = {Radiative processes in astrophysics},
	url = {https://ui.adsabs.harvard.edu/abs/1979rpa..book.....R},
	urldate = {2023-08-17},
	author = {Rybicki, George B. and Lightman, Alan P.},
	month = jan,
	year = {1979},
	note = {Publication Title: A Wiley-Interscience Publication
ADS Bibcode: 1979rpa..book.....R},
}

@article{blandford_fluid_1976,
	title = {Fluid dynamics of relativistic blast waves},
	volume = {19},
	issn = {0031-9171},
	url = {https://aip.scitation.org/doi/abs/10.1063/1.861619},
	doi = {10.1063/1.861619},
	number = {8},
	urldate = {2022-09-21},
	journal = {The Physics of Fluids},
	publisher = {American Institute of Physics},
	author = {Blandford, R. D. and McKee, C. F.},
	month = aug,
	year = {1976},
	pages = {1130--1138},
}

@article{waxman_angular_1997,
	title = {Angular {Size} and {Emission} {Timescales} of {Relativistic} {Fireballs}},
	volume = {491},
	issn = {0004-637X},
	url = {https://doi.org/10.1086/311057},
	doi = {10.1086/311057},
	language = {en},
	number = {1},
	urldate = {2022-09-21},
	journal = {The Astrophysical Journal},
	publisher = {American Astronomical Society},
	author = {Waxman, Eli},
	year = {1997},
	pages = {L19--L22},
}

@article{nakar_short-hard_2007,
	series = {The {Hans} {Bethe} {Centennial} {Volume} 1906-2006},
	title = {Short-hard gamma-ray bursts},
	volume = {442},
	issn = {0370-1573},
	url = {https://www.sciencedirect.com/science/article/pii/S0370157307000476},
	doi = {10.1016/j.physrep.2007.02.005},
	language = {en},
	number = {1},
	urldate = {2022-09-20},
	journal = {Physics Reports},
	author = {Nakar, Ehud},
	month = apr,
	year = {2007},
	pages = {166--236},
}

@article{sari_spectra_1998,
	title = {Spectra and {Light} {Curves} of {Gamma}-{Ray} {Burst} {Afterglows}},
	volume = {497},
	issn = {0004-637X},
	url = {https://iopscience.iop.org/article/10.1086/311269/meta},
	doi = {10.1086/311269},
	language = {en},
	number = {1},
	urldate = {2022-09-20},
	journal = {The Astrophysical Journal},
	publisher = {IOP Publishing},
	author = {Sari, Re'em and Piran, Tsvi and Narayan, Ramesh},
	month = mar,
	year = {1998},
	pages = {L17},
}

@article{waxman_gamma-ray_2006,
	title = {Gamma-ray bursts and collisionless shocks},
	volume = {48},
	issn = {0741-3335},
	url = {https://doi.org/10.1088/0741-3335/48/12b/s14},
	doi = {10.1088/0741-3335/48/12B/S14},
	language = {en},
	number = {12B},
	urldate = {2022-09-20},
	journal = {Plasma Physics and Controlled Fusion},
	publisher = {IOP Publishing},
	author = {Waxman, E.},
	year = {2006},
	pages = {B137--B151},
}

@article{bykov_fundamentals_2011,
	title = {Fundamentals of collisionless shocks for astrophysical application, 2. {Relativistic} shocks},
	volume = {19},
	issn = {1432-0754},
	url = {https://doi.org/10.1007/s00159-011-0042-8},
	doi = {10.1007/s00159-011-0042-8},
	language = {en},
	number = {1},
	urldate = {2022-09-20},
	journal = {The Astronomy and Astrophysics Review},
	author = {Bykov, A. M. and Treumann, R. A.},
	month = aug,
	year = {2011},
	pages = {42},
}

@article{piran_physics_2005,
	title = {The physics of gamma-ray bursts},
	volume = {76},
	url = {https://link.aps.org/doi/10.1103/RevModPhys.76.1143},
	doi = {10.1103/RevModPhys.76.1143},
	number = {4},
	urldate = {2022-09-20},
	journal = {Reviews of Modern Physics},
	publisher = {American Physical Society},
	author = {Piran, Tsvi},
	month = jan,
	year = {2005},
	pages = {1143--1210},
}

@article{blandford_particle_1987,
	title = {Particle acceleration at astrophysical shocks: {A} theory of cosmic ray origin},
	volume = {154},
	issn = {0370-1573},
	shorttitle = {Particle acceleration at astrophysical shocks},
	url = {https://www.sciencedirect.com/science/article/pii/0370157387901347},
	doi = {10.1016/0370-1573(87)90134-7},
	language = {en},
	number = {1},
	urldate = {2022-09-20},
	journal = {Physics Reports},
	author = {Blandford, Roger and Eichler, David},
	month = oct,
	year = {1987},
	pages = {1--75},
}

@article{sironi_maximum_2013,
	title = {The maximum energy of accelerated particles in relativistic collisionless shocks},
	volume = {771},
	issn = {15384357},
	doi = {10.1088/0004-637X/771/1/54},
	number = {1},
	journal = {Astrophysical Journal},
	publisher = {Institute of Physics Publishing},
	author = {Sironi, Lorenzo and Spitkovsky, Anatoly and Arons, Jonathan},
	month = jul,
	year = {2013},
}

@article{keshet_energy_2005,
	title = {Energy spectrum of particles accelerated in relativistic collisionless shocks},
	volume = {94},
	issn = {00319007},
	doi = {10.1103/PhysRevLett.94.111102},
	number = {11},
	journal = {Physical Review Letters},
	author = {Keshet, Uri and Waxman, Eli},
	month = mar,
	year = {2005},
}
\appendix
\section{Self-absorption frequencies}
\label{app:self}
We estimate the self-absorption frequency at time $t$ as the frequency for which the optical depth is $\tau_\nu=\alpha_\nu\Delta_\tau=1$, where $\alpha_\nu$ and $\Delta_\tau$ are the typical absorption coefficient and the typical path length traversed by photons through the shocked plasma, dominating the emission of radiation observed at time $t$.

To derive the self-absorption frequency in the observer frame, we first approximate the absorption coefficient \citep{rybicki_radiative_1979}.
For the FS $\nu_a\ll\nu_m$,
\begin{equation}
    \alpha^f_\nu = g^f_a(p)\rho\left(\gamma_m^f\right)^{-\frac{5}{3}}(1+z)^{-\frac{2}{3}}\left(\gamma^f\right)^{\frac{5}{3}}\nu^{-\frac{5}{3}}\left(B^f\right)^\frac{2}{3},
\end{equation}
where
\begin{equation}
g_a^f(p)=\frac{(p+2)(p-1)}{(3p+2)}\frac{12e^3\int_0^\pi\sin^\frac{5}{3}(y)dy}{2^\frac{7}{3}m_pm_e^2c^2\Gamma\left(\frac{1}{3}\right)}\left[\frac{4\pi m_ec}{3e}\right]^{\frac{1}{3}}.
\end{equation}
For the RS $\nu_m\ll\nu_a$,
\begin{equation}
    \alpha^r_\nu = g^r_a(s,p)\rho_\text{ej}\left(\gamma_m^r\right)^{p-1}(1+z)^{-\frac{p+2}{2}}\left(\gamma^f\right)^\frac{p+2}{2}\nu^{-\frac{p+4}{2}}\left(B^r\right)^\frac{p+2}{2},
\end{equation}
where
\begin{equation}
\begin{aligned}
g_a^r(s,p)=2\left(x+\frac{1}{x}\right)\frac{(p-1)\sqrt{3}e^3}{16\sqrt{\pi}m_pm_e^2c^2}\left[\frac{3e}{2\pi m_ec}\right]^{\frac{p}{2}}\times\\\frac{\Gamma\left(\frac{p+6}{4}\right)}{\Gamma\left(\frac{p+8}{4}\right)}\Gamma\left(\frac{3p+22}{12}\right)\Gamma\left(\frac{3p+2}{12}\right).
\end{aligned}
\end{equation}
The photons are emitted in the shocked plasma while it expands relativistically with LF $\gamma_f$. The typical distance they go through in the shocked external medium is
\begin{align}
\Delta_\tau^f\approx\Delta_\text{ext}\gamma_f^2,
\end{align}
and in the shocked ejecta
\begin{align}
\Delta_\tau^r\approx\Delta_\text{ej}\gamma_f^2.
\end{align}
The self-absorption frequency, $\nu_{a}$, is simply defined by  $\tau_\nu(\nu=\nu_{a})=1$. For the relevant parameter space, $\nu_a^r\gg\nu_a^f$. Thus, even though the photons emitted from the shocked ejecta go through the shocked external medium, they are not affected by its self-absorption.
\section{GENERAL SYNCHROTRON SCALINGS}
\label{app:k}
Here, we summarize the temporal scalings of the synchrotron observables for a general external-density profile $\rho_1=A r^{-k}$, with $k<3$, during the phase in which the reverse shock propagates through the stratified ejecta. The temporal scaling of the FS LF and radius are
\begin{equation}
\gamma_f\propto t^{-\frac{3-k}{s+7-2k}},\qquad R \propto t^\frac{s+1}{s+7-2k}
\end{equation}
Using Eqs.~(\ref{eq:gammam})--(\ref{eq:nusyn}) and the same assumptions adopted in the main text, the characteristic synchrotron quantities scale with observer time as follows. For the FS,
\begin{align}
\nu_m^f &\propto t^\frac{7k-ks-24}{2(s+7-2k)}, \\
\nu_c^f &\propto t^\frac{(3k-4)(s+1)}{2(s+7-2k)}, \\
F_{\nu,\max}^f &\propto t^\frac{(k-3ks+6s-6)}{2(s+7-2k)} .
\end{align}
For the RS,
\begin{align}
\nu_m^r &\propto t^\frac{3k-ks-12}{2(s+7-2k)}, \\
\nu_c^r &\propto t^\frac{(3k-4)(s+1)}{2(s+7-2k)}, \\
F_{\nu,\max}^r &\propto t^\frac{3(k-ks+2s-4)}{2(s+7-2k)} .
\end{align}
For the self-absorption frequencies in the spectral ordering relevant here, $\nu_a^f<\nu_m^f<\nu_c^f$ and $\nu_m^r<\nu_a^r<\nu_c^r$, we obtain
\begin{align}
\nu_a^f &\propto t^\frac{3s-4ks-2k-3}{5(s+7-2k)}, \\
\nu_a^r &\propto t^\frac{3kp-kps-6ks+10k-12p+4s-44}{2(p+4)(s+7-2k)}.
\end{align}
\bsp	
\label{lastpage}
\end{document}